%% file: ms.tex
\pdfoutput=1

\documentclass{IEEEtran} 
\usepackage{mathptmx} 

\usepackage{graphicx}
\newcommand{\ignore}[1]{}
\usepackage{fancyhdr}
\usepackage[normalem]{ulem}
\usepackage[hyphens]{url}

\usepackage[sort,nocompress]{cite}
\usepackage{microtype}

\usepackage[bookmarks=false,letterpaper=true,colorlinks=true,linkcolor=black,citecolor=blue,urlcolor=black]{hyperref}

\usepackage{alltt}
\usepackage{epsfig}
\usepackage{graphicx}
\usepackage{enumitem}
\usepackage{amsmath}

\usepackage{amssymb}
\usepackage[normalem]{ulem}
\usepackage{float}
\usepackage[]{algorithm2e}
\usepackage{lipsum}
\usepackage{balance}
\usepackage{tikz}
\usepackage{calc}
\usepackage{xcolor, colortbl}
\usepackage{multirow}
\usepackage{pifont}
\usepackage{adjustbox}
\usepackage{array}

\usepackage{xfrac}

\ignore{
\usepackage{color,soul}
\definecolor{lightgreen}{RGB}{195, 233, 211}
\sethlcolor{lightgreen}

\soulregister\cite7
\soulregister\ref7
\soulregister\pageref7

\makeatletter
\def\SOUL@hlpreamble{%
	\setul{}{2.2ex}
	\let\SOUL@stcolor\SOUL@hlcolor
	\SOUL@stpreamble
}
\makeatother
}

\usepackage{amsfonts}
\usepackage{url}

\ignore{
\definecolor{Green}{rgb}{0, 1, 0}
\definecolor{Red}{rgb}{1, 0, 0}
\newcolumntype{R}[2]{%
    >{\adjustbox{angle=#1,lap=\width-(#2)}\bgroup}%
    l%
    <{\egroup}%
}

\newcolumntype{g}{>{\columncolor{Green}}c}
\newcolumntype{r}{>{\columncolor{Red}}c}

\setlength\extrarowheight{1pt}
}

\fancypagestyle{firstpage}{
	\fancyhf{}
	\setlength{\headheight}{0pt}
	
	\pagenumbering{arabic}
}

\title{
\vspace{-0.1in}
CRAM: Efficient Hardware-Based Memory Compression for Bandwidth Enhancement
}

\usepackage{authblk}
\author[ ]{Vinson Young}
\author[ ]{Sanjay Kariyappa}
\author[ ]{Moinuddin K. Qureshi\vspace{-.1in}}
\affil[ ]{Georgia Institute of Technology}
\affil[ ]{\texttt {\{vyoung,sanjaykariyappa,moin\}@gatech.edu}\vspace{-.2in}}

\begin{document}
\maketitle
\thispagestyle{firstpage}
\pagestyle{plain}
\input{abstract}

\vspace{-0.23in}

\input{introduction}

\newpage
\input{motivation}
\input{methodology}

\input{organization}

\input{results}


\input{related}

\input{summary}
\bibliographystyle{IEEEtran.bst}
\bibliography{references}

\end{document}

%% file: abstract.tex
\ignore{
FLOW:
1. What we are targeting - compression for bandwidth..

2. What are the key challenges with compression? location

3. How do other works address these challenges? page table, metadata cache

4. How we address these challenges? predict, signature, toggling compression by sampling

}


\begin{abstract}


This paper investigates hardware-based memory compression designs to increase the memory bandwidth.  When lines are compressible, the hardware can store multiple lines in a single memory location, and retrieve all these lines in a single access, thereby increasing the effective memory bandwidth.  However, relocating and packing multiple lines together depending on the compressibility causes a line to have multiple possible locations.  Therefore, memory compression designs typically require metadata to specify the compressibility of the line.  Unfortunately, even in the presence of dedicated metadata caches, maintaining and accessing this metadata incurs significant bandwidth overheads and can degrade performance by as much as 40\%. Ideally, we want to implement memory compression while eliminating the bandwidth overheads of  metadata accesses.

This paper proposes {\em CRAM}, a bandwidth-efficient design for  memory compression that is entirely hardware based and does not require any OS support or changes to the memory modules or interfaces. CRAM uses a novel {\em implicit-metadata} mechanism, whereby the compressibility of the line can be determined by scanning the line for a special {\em marker} word,  eliminating the overheads of metadata access. CRAM is equipped with a low-cost {\em Line Location Predictor (LLP)} that can determine the location of the line  with 98\% accuracy. Furthermore, we also develop a scheme that can dynamically enable or disable compression based on the bandwidth cost of storing compressed lines and the bandwidth benefits of obtaining compressed lines, ensuring no degradation for workloads that do not benefit from compression.  Our evaluations, over a diverse set of 27 workloads, show that CRAM  provides a speedup of up to 73\% (average 6\%) without causing slowdown for any of the workloads, and consuming a storage overhead of less than 300 bytes at the memory controller. 

\end{abstract}


\ignore{



By storing cachelines in a compressed form, we can fetch more than one cacheline per memory request, enabling a higher effective memory bandwidth. Packing compressed cachelines together to achieve greater bandwidth comes at the cost of variable location of cachelines as a function of compressibility. The key challenge therefore is to determine the location of potentially compressed cachelines. Prior works on memory compression try to solve this problem by having some form of metadata cache to store the compressed location of cachelines. Additional metadata lookup is a bandwidth overhead for workloads with low spatial locality. This ends up hurting performance significantly if the cachelines are incompressible.



We solve this problem by eliminating the need to do metadata lookup altogether. We propose a 'tagged compression' scheme, which involves appending a special tag to compressed cachelines. This helps identify compressed cachelines without consulting a metadata table. We design a compressibility predictor that can be used with this scheme to predict the location of cachelines. The compression tag provides a way to confirm our prediction with almost no overheads. By eliminating the additional accesses required for fetching metadata, we avoid degrading the performance for incompressible workloads while still enjoying the benefits of compression for compression friendly workloads.

We also find through our studies that compression can cause performance degradation for  workloads which have low data reuse. This is because of the bandwidth overheads of writing back compressed cachelines to memory.
We suggest a simple sampling scheme to dynamically turn off compression when the costs outweigh the benefits. This eliminates the overheads associated with compression altogether when it is not beneficial. 
}

%% file: introduction.tex
\ignore{
FLOW:
1. Bandwidth wall
2. How compression can help with figure

}

\section{Introduction}

As modern compute systems pack more and more cores on the processor chip, the memory systems must also scale proportionally in terms of bandwidth in order to supply data to all the cores. Unfortunately, memory bandwidth is dictated by the pin count of the processor chip, and this limited memory bandwidth is one of the bottlenecks for system performance.  Data compression is a promising solution for enabling a higher effective memory bandwidth. For example, when the data is compressible, we can pack multiple memory lines within one memory location and retrieve all of these lines with a single memory request, increasing memory bandwidth and performance.  In this paper, we study main memory compression designs that can increase memory bandwidth.


Prior works on memory compression~\cite{IBM-MXT:HPCA2001}\cite{Strenstorm:ISCA2005}\cite{LCP:MICRO2013}  aim to obtain both the capacity and bandwidth benefits from compression, trying to accommodate as many pages as possible in the main memory, depending on the compressibility of the data. As the effective memory capacity of such designs can change at runtime, these designs need support from the Operating System (OS) or the hypervisor, to handle the dynamically changing memory capacity.  Unfortunately, this means such memory compression solutions are not viable unless both the hardware vendors (e.g. Intel, AMD etc.) and the OS vendors (Microsoft, Linux etc.) can co-ordinate with each other on the interfaces, or such solutions will be limited to systems where the same vendor provides both the hardware and the OS. We are interested in practical designs for memory compression that can be implemented entirely in hardware, without relying on any OS/hypervisor support. Such designs would provide the bandwidth benefits, while providing constant memory capacity.\footnote{In fact, a few months ago, Qualcomm's Centriq\cite{qualcomm} system was announced with a feature that tries to provide higher bandwidth through memory compression while forgoing the extra capacity available from memory compression. Centriq's design relies on increasing the linesize to 128 bytes, striping this wider line across two channels, having ECC DIMMs in each channel to track compression status, and obtaining the 128-byte line from one channel if the line is compressible. Ideally, we want to obtain bandwidth benefits without changing the linesize, or relying on ECC DIMMs, and without getting limited to 2x compression ratio. Nonetheless, the Centriq announcement shows the commercial appeal of such hardware-based memory compresssion. }

\begin{figure*}[htb]
	\centering
    \vspace{-0.15 in}
\centerline{\epsfig{file=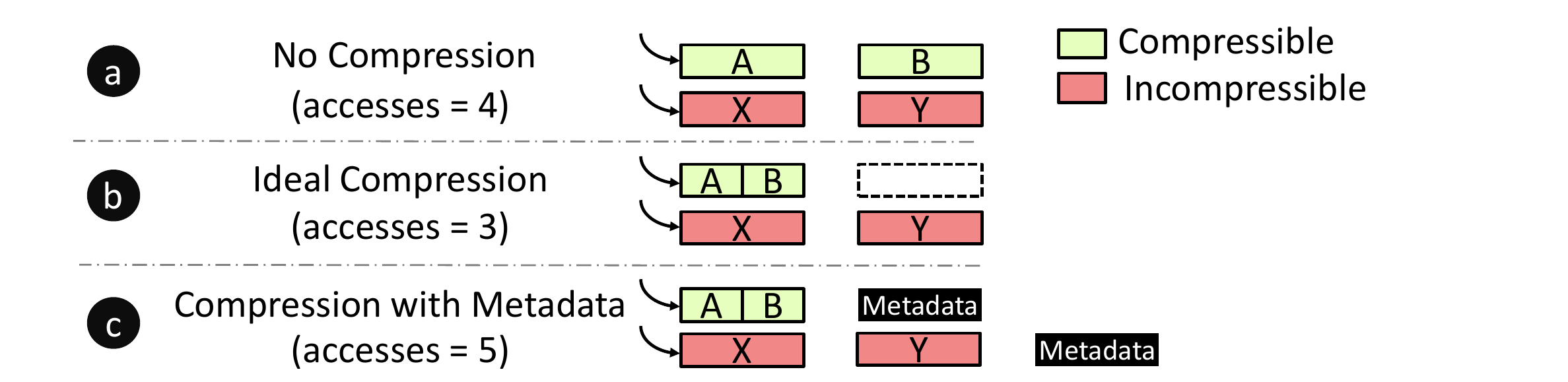, height = 1.5 in}}
    \vspace{.06 in}
	\caption{Number of accesses to read 4 lines: A, B, X, Y for (a) uncompressed memory, (b) ideal compressed memory, and (c) compressed memory with metadata lookup overhead. Metadata lookup causes significant bandwidth overhead.}
	\vspace{-0.06 in}
    
    \label{fig:intro}
\end{figure*}

A prior study, MemZip~\cite{shafiee2014memzip}, tried to increase the memory bandwidth using hardware-based compression; however, it requires significant changes to the memory organization and the memory access protocols. Instead of striping the line across all the chips on a memory DIMM, MemZip places the entire line in one chip, and changes the number of bursts required to stream out the line, depending on the compressibility of the data.  Thus, MemZip  requires significant changes to the data organization of commodity memories and the memory controller to support variable burst lengths. Ideally, we want to obtain the memory bandwidth benefits from compression while retaining support for commodity DRAM modules and using conventional data organization and bus protocols.

  \ignore{
Furthermore, all prior proposals on memory compression, require per-line metadata to indicate the {\em Compressibility Status Information (CSI)} of the line, and this information is used to determine the location of the line (or the burst length for MemZip).  Even in the presence of dedicated caches,  obtaining this metadata can incur significant bandwidth overheads. 
}

Compression can change both the size and location of the line.  Without additional information, the memory controller would not know how to interpret the data  obtained from the memory (compressed or uncompressed). Conventional designs for memory compression rely on explicit metadata that indicates the compressibility status of the line, and  this information is used to determine the location of the line.  Such designs store the  metadata in a separate region of memory. Unfortunately, accessing the metadata can incur significant bandwidth overhead. While on-chip metadata caches~\cite{LCP:MICRO2013} reduce the bandwidth required to obtain the metadata, such caches are designed mainly to exploit spatial locality and are not as effective for workloads that have limited spatial locality.  Our goal is to design hardware-compressed memory, without any OS support, using commodity memories and protocols, and without the metadata lookup.



\ignore{
\begin{table}[hbt]
	\centering
     \vspace{-0.1 in}
    \begin{small}
	\caption{Comparing Main-Memory Compression Design}
  \setlength{\tabcolsep}{0.6em}
	\begin{tabular}{|c||c|c|c|}\hline
		& OS Support & Changes to Memory       & Metadata\\ 
        & Needed ?   & Organization/Protocol?  & Lookup? \\ \hline 
        
        LCP~\cite{LCP:MICRO2013}& Yes & {\bf  No} & Yes \\ \hline 
	    MemZip~\cite{shafiee2014memzip}& {\bf No} & Yes & Yes \\ \hline 
        Our Goal & {\bf No} & {\bf No} & {\bf No} \\ \hline    
	\end{tabular}
	\label{tab:intro}
    \vspace{-0.15 in}
	\end{small}
\end{table}
}

\newpage

We explain the problem of bandwidth overhead of metadata with an example. Figure~\ref{fig:intro} shows three memory systems, each servicing four memory requests A, B, X and Y.  A and B are compressible and can reside in one line, whereas X and Y are incompressible. For the baseline system (a), servicing these four requests would require four memory accesses.  For an idealized compressed memory system (b) (that does not require metadata lookup), lines A  and B can be obtained in a single access, where as X and Y would require one access each, for a total of 3 accesses for all the four lines.  However, when we account for metadata lookup (c), it could take up to 5 accesses to read and interpret all the lines, causing degradation relative to an uncompressed scheme. Our studies show that even in the presence of metadata caching, the metadata lookup can degrade performance by as much as 40\%. Ideally, we want to implement memory compression without incurring the bandwidth overheads of metadata accesses.

To this end, this paper presents {\em CRAM}, an efficient hardware-based main-memory compression design. CRAM decouples and separately solves the issue of (i) how to interpret the data, and (ii) where to look for the data, to  eliminate the  metadata lookup. To efficiently interpret the data received on an access, we propose an \emph{implicit-metadata} scheme, whereby compressed lines are required to contain a special value, called a {\em marker}. For example, with a four-byte marker, the last four bytes of a compressed line is required to always be equal to the marker value.  We leverage the insight that compressed data rarely uses the full 64-byte space, so we can store compressed data within 60 bytes and  use the remaining four bytes to store the marker.  On a read to a line that contains the marker, the line is interpreted as a compressed line. Similarly, an access to a line that does not contain the marker is interpreted as an uncompressed line.  The likelihood that an uncompressed line coincidentally matches with a  marker is quite small (less than one in a billion), and CRAM handles such rare cases of marker collisions simply by identifying lines that cause marker collisions on a write and storing such lines in an inverted form  (more details in Section~\ref{label:collisions}).

The implicit-metadata scheme eliminates the need to do a separate metadata lookup. However,  CRAM now needs an efficient mechanism to determine the location of the given line. CRAM restricts the possible locations of the line, based on compressibility. For example, in Figure~\ref{fig:intro}(b), when A and B are compressible, CRAM restricts that both A and B must reside in the location of A.  Therefore, the location of A remains unchanged regardless of compression.  However, for B, the location depends on compressibility. We propose a history-based \textit{Line Location Predictor (LLP)}, that can identify the correct location of the line with a high accuracy, which helps in obtaining a given line in a single memory access. The LLP is based on the observation that lines within a page tend to have similar compressibility. We propose a page-based last-compressibility predictor to predict compressibility and thus location, and this allows us to access the correct location with 98\% accuracy. CRAM, combined with   implicit-metadata and LLP, eliminates metadata lookups and achieves an average speedup of 8.5\% on spec workloads.


Unfortunately, even after eliminating the bandwidth overheads of the metadata lookup, some workloads still have slowdown with compression due to the inherent bandwidth overheads associated with compressing memory. For
example, compressing and writing back clean-lines incurs bandwidth overhead, as those lines are not written to memory in an uncompressed design. For workloads with poor reuse, this bandwidth overhead of writing compressed data does not get amortized by the subsequent accesses. To avoid performance degradation in such scenarios, we develop {\em Dynamic-CRAM}, a sampling-based scheme that can dynamically enable or disable compression depending on when compression is beneficial. Dynamic-CRAM ensures no slowdown for workloads that do not benefit from compression.

\vspace{0.05 in}

\noindent Overall, this paper makes the following contributions:
\vspace{0.05 in}



\noindent 1. It proposes CRAM, a hardware-based compressed memory design  to provide bandwidth benefits without requiring OS support or changes to the memory module and protocols. CRAM performs memory accesses using conventional linesize (64 bytes) and does not rely on the availability of ECC-DIMMs.

\vspace{0.05 in}

\noindent 2. It proposes {\em implicit-metadata} to eliminate the storage and bandwidth overheads of metadata, by providing both the compressibility status and data in a single memory access.

\vspace{0.05 in}

\noindent 3. It proposes a low-cost {\em Line Location Predictor (LLP)} to determine the location of the line with a high accuracy.
\vspace{0.05 in}

\noindent 4. It proposes Dynamic-CRAM, to enable or disable compression at runtime based on the cost and benefit of compression.

\vspace{.05in}

Our evaluations show that CRAM improves bandwidth by 9\% and provides a speedup of up to 73\% (average 6\%), while ensuring no slowdown for any of the workloads.  CRAM can be implemented with minor changes to the memory controller, while incurring a storage overhead of less than 300 bytes.

\ignore{

\ignore{
    Memzip~\cite{shafiee2014memzip} compresses data in-place and uses smaller burst-lengths to try to achieve bandwidth benefits. Memzip explicitly looks up metadata to learn compressibility and burst-length, and it sacrifices capacity benefits to maintain OS-transparency. However, it needs non-commodity DRAM to achieve its bandwidth benefits. We take its approach for OS transparency--we sacrifice capacity benefits in order to retain OS transparency.

	LCP uses packs multiple contiguous lines together to get bandwidth benefits on commodity DRAM. However, it requires OS support and metadata lookup to interpret the lines. We take a similar approach and pack together multiple lines so that our scheme can work on commodity DRAM.
    
	However, we must solve the key flaw of metadata lookup if we are to maintain robustness when workload footprints scale and metadata cache hit-rates become poor.

Several recent studies have looked at using compression for main memories~\cite{shafiee2014memzip,LCP:MICRO2013}. Such works pack data closely, so that they can retrieve data from memory in a fewer number of bursts or accesses.
However, such works suffer from two major problems: they require a cross-layer approach that can limit \textit{practicality} for adoption, and they require \textit{metadata lookups} that can incur significant bandwidth overheads.

In terms of \textit{practicality} for adoption, prior works either require changes to the Operating System (OS) or require employing specialized memory modules.

to multiple layers of the computing stack to enable their solution. We show an overview in Table~\ref{tab:motivation_table}.
First, prior techniques often rely specialized DRAM DIMMs to implement their ideas.

This limits its viability for adoption in the near future. In our work, we obviate the need for specialized DIMMs by choosing to instead store multiple contiguous lines together and stream them out in one access, to achieve bandwidth benefits while maintaining applicability to commodity DRAM.

For example, 
}

}

%% file: motivation.tex
\ignore{
FLOW:
- Overview of compressed memory
- Elements (metadata, location, clean writes)
- Potential improvement
- The Problem of Metadata Access
- Insight and Solution
}

\section{Background and Motivation}

Compression exploits redundancy in data values and can provide both larger effective capacity and higher effective bandwidth for the memory systems.  While exploiting the increased memory capacity requires OS support (to handle the dynamically changing capacity depending on data values), memory compression for exploiting only the bandwidth benefits can potentially be implemented entirely in hardware.  We provide background on hardware-based memory compression, the potential benefit from compression, the challenges in implementing such a compressed design, and insights that can help in developing efficient designs.

\ignore{		
Why HW Compression for BW. 2x channel. OS-invisible
		Problem: metadata lookup. Need figure (bw overhead to access? Perf degradation)
		Insight: store compression status bit. (instead of 64B, store 60B. Same 
			Figure. 
}

\subsection{Memory Compression for Bandwidth}

\begin{figure}[htb]
\vspace{-0.08 in}
	\centering
 \centerline{\epsfig{file=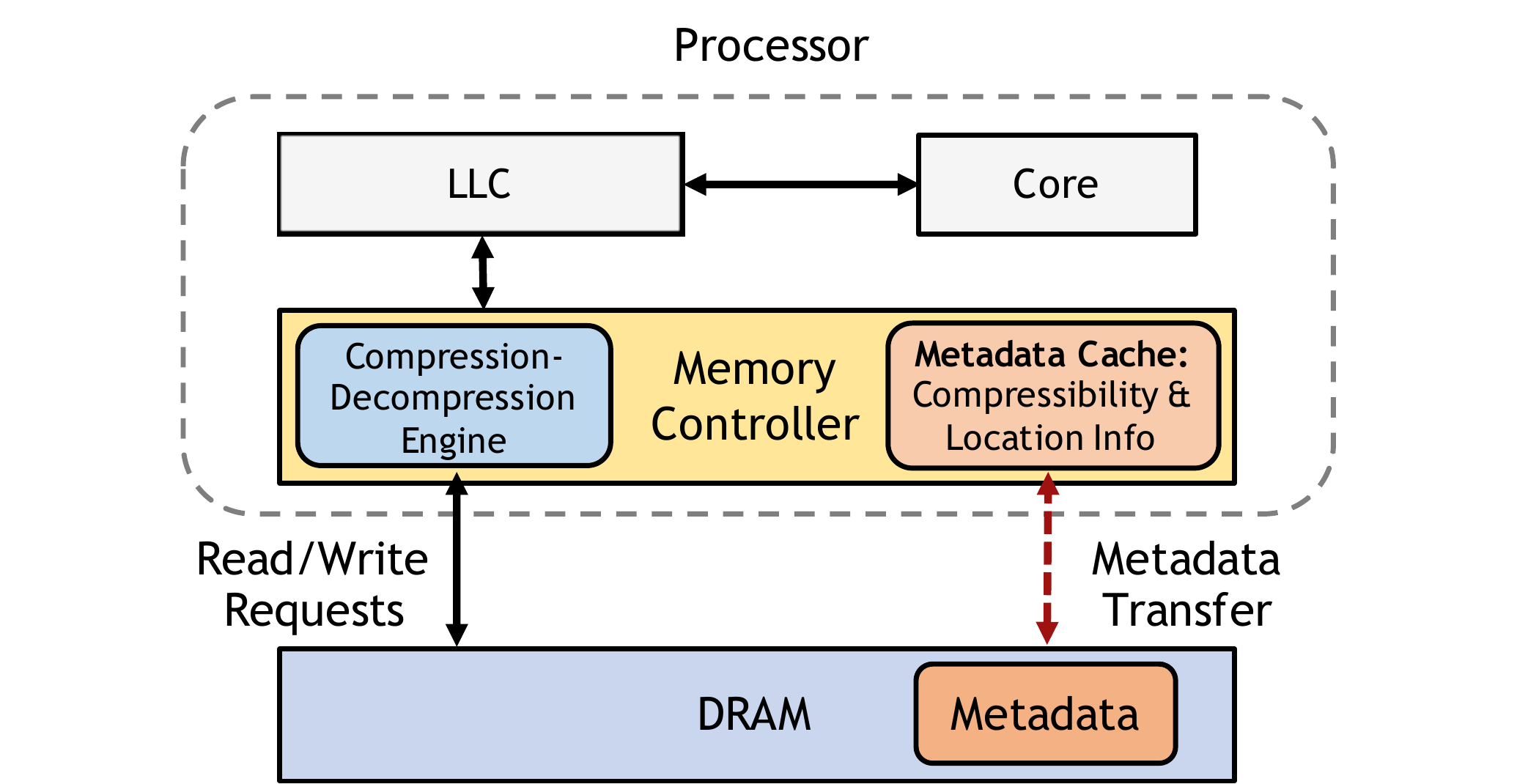, width=3.5in}}
	\caption{Overview of Compressed Memories. Conventional designs track compression status of each line in a metadata region and use a metadata cache.}
	\vspace{-0.20 in}
	\label{fig:arch}
\end{figure}

Figure~\ref{fig:arch} provides an overview of a compressed memory design.  We will assume that the design is geared towards obtaining only the bandwidth benefits, and the extra capacity created by compression  remains unused.  Compressed memory designs leverage compression algorithms\cite{FPC}\cite{zca}\cite{zhang2000frequent}\cite{fvc}\cite{BDI}\cite{Alaa:ISCA2004}\cite{bitplane} to accommodate data into a smaller space. If the lines are compressible we can either store them in their original location and stream out in a smaller burst, or place multiple compressed lines in one location and stream out all these lines in one access.   We use the second option as it avoids dynamically changing the burst length, thus retaining compliance with the protocols and data mapping used in conventional memory designs.  Thus, if two lines A and B are compressible, then both are resident in the location of A.  One memory access can provide both A and B, thereby increasing the effective bandwidth if both A and B get used.
The bandwidth benefit of such a design is dictated by both (a) the ability to pack multiple lines together, and (b) the spatial locality of the workload (ability to use adjacent lines).

\subsection{The Challenge of Metadata Accesses}

An access to the compressed memory obtains a 64-byte line, however, the memory controller would not know if the line contains compressed data or not. For example, an access to A would provide both A and B, if both lines are compressible, and only A if the lines are uncompressed.  Simply obtaining the line from location A is insufficient to provide the information about compressibility of the line.  A separate region in memory, which we refer to as the {\em metadata} region keeps track of the {\em Compression Status Information (CSI)} for each line.  Thus, we need the CSI of the line along with the data line to not only interpret the data line, but also to determine the location of the data line. Even if we provisioned only one bit per line in memory, the size of this metadata would be quite large. For example, for our 16GB memory, having 1-bit per line to specify if the line is compressed or not would require a capacity of 32 megabytes.  Therefore, conventional designs keep the metadata region in memory and access this metadata region on a demand basis and cache it in an on-chip metadata cache~\cite{LCP:MICRO2013}\cite{shafiee2014memzip}.  Such designs are effective only when the metadata cache has a high hit-rate, due to either high spatial locality or small workload footprint. However, these approaches become ineffective when scaled to much larger workloads with low spatial locality. In the worst-case, these designs may need a separate metadata access for every data access, constituting a potential bandwidth overhead of 50-80\% (e.g., in \textit{xz} and \textit{cactu}). Therefore, avoiding the bandwidth overhead of metadata accesses is vital to building an effective  memory compression design.

\subsection{Potential for Performance Improvement}

\begin{figure}[htb]
	\centering
    \vspace{-.1in}
\centerline{\epsfig{file=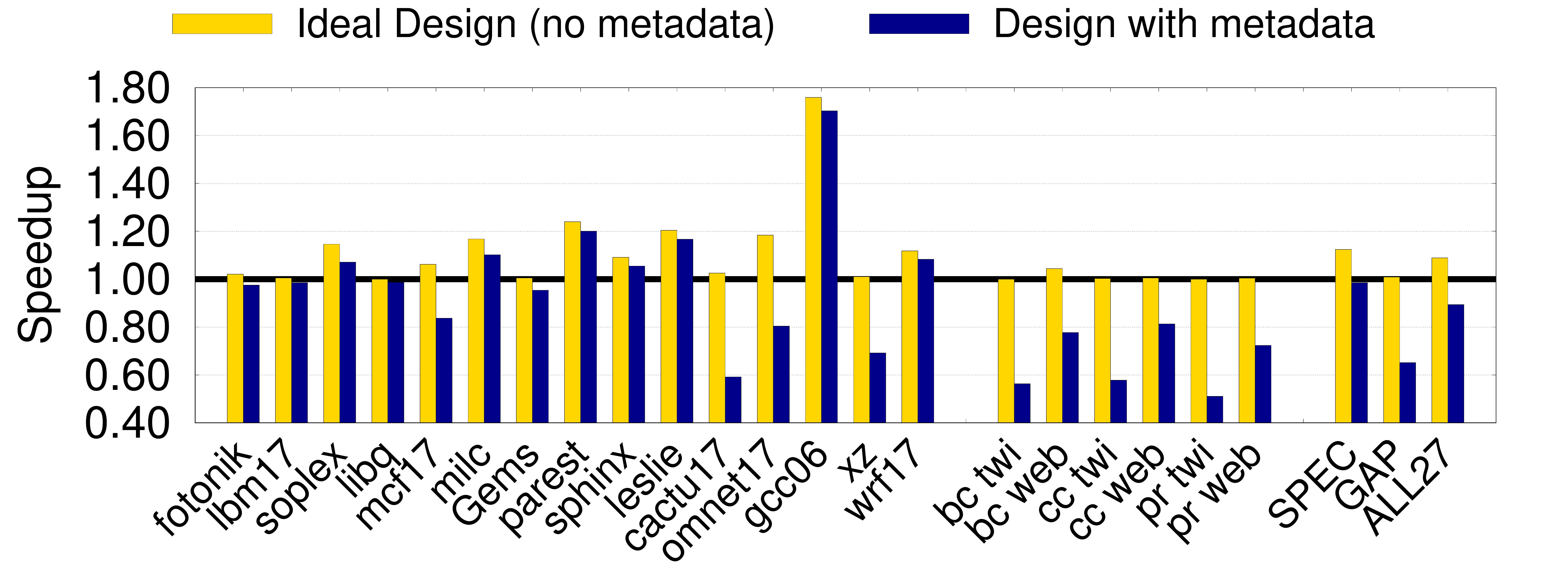, width=\columnwidth}}

    \vspace{-.13in}
	\caption{Speedup from ideal compression (no metadata lookup) and practical compression (w/ metadata cache).}
    \vspace{-.1in}
	\label{fig:ideal}
\end{figure}

Our goal is to develop an efficient memory compression design that provides higher bandwidth.  Figure~\ref{fig:ideal} shows the performance benefit from an idealized compression scheme that does not maintain any metadata and simply transfers all the lines that would be together in a compressed memory system, thereby obtaining all the benefits of compression and none of the overheads. We also show the performance of a practical memory compression that maintains metadata in memory and is equipped with a 32KB metadata cache.

On average, compression can provide a speedup of 9\%; however, the overheads of implementing compression  erodes this.  In fact, we observe significant degradation with compression for several workloads.  For example, Graph workloads (bc\_twi - pr\_web) have small potential benefits from compression, due to poor spatial locality and low data reuse. It is important that the implementation of compressed memory does not degrade the performance of such workloads.

\subsection{Insight:  Store Metadata in Unused Space}

\begin{figure}[htb]
	\centering
\centerline{\epsfig{file=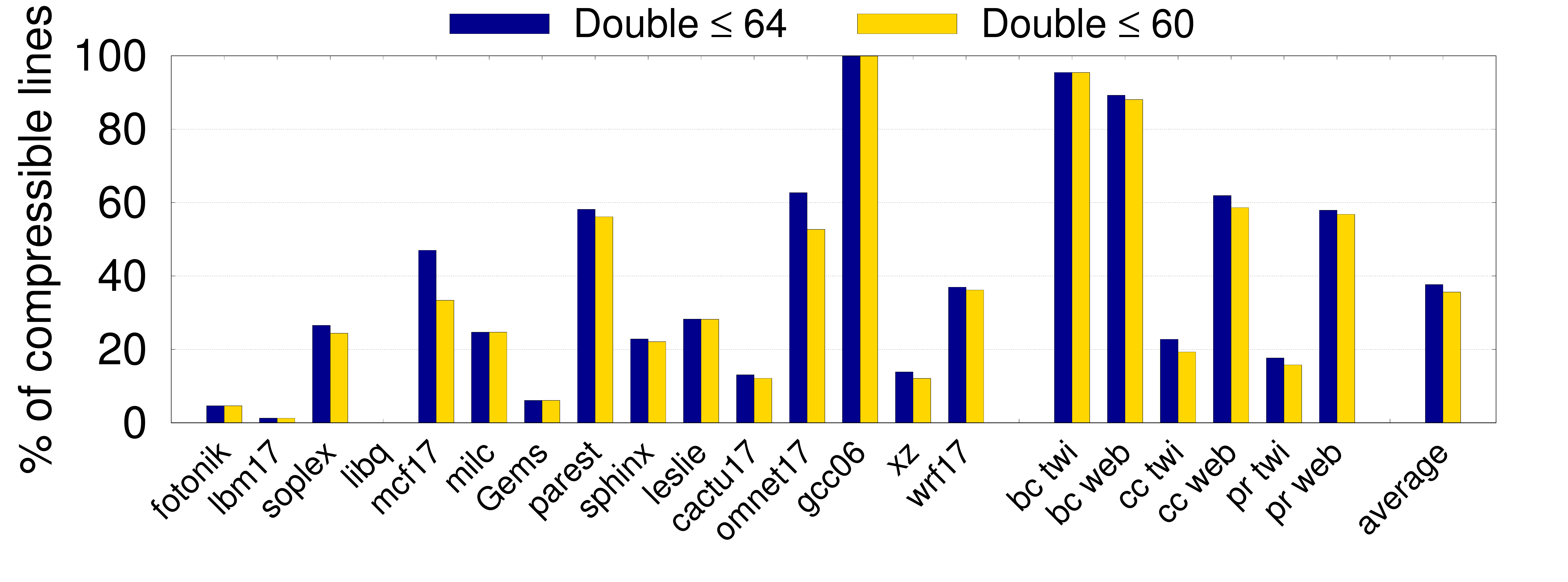, width=\columnwidth}}
	\vspace{-0.08 in}
	\caption{Probability of a pair of adjacent lines compressing to $\leq$64B and $\leq$60B. We can use 4 bytes to indicate   status of the line, without significantly affecting compressibility.}
	\label{fig:motivation_insight}
\end{figure}

To reduce the metadata access of compressed memory, we leverage the insight that not all the space of the 64 byte line is used by compressed memory.  For example, when we are trying to compress two lines (A and B) they must fit within 64 bytes; however, the compressed size could still be smaller than 64 bytes (and not large enough to store additional lines C and D).  We can leverage the unused  space in the compressed memory line to store metadata information within the line. For example, we could require that the compressed lines store a 4-byte marker (a predefined value) at the end of the line, and the space available to store the compressed lines would now get reduced to 60 bytes.  Figure~\ref{fig:motivation_insight} shows the probability of a  pair of adjacent lines compressing to $\leq$64B and $\leq$60B. As the probability of compressing pairs of lines to $\leq$64B and $\leq$60B are 38\% and 36\%, respectively, we find that reserving space for this marker does not substantially impact the likelihood of compressing two lines together and thus would not have a significant impact on compression ratio.

We can use this insight to store the metadata implicitly within the line and avoid the bandwidth overheads of accessing the metadata explicitly.  If the line obtained from memory contains the marker value, the line is deemed compressed, whereas, if the the line does not have the marker value, then it is deemed uncompressed.  However, there could be a case where the uncompressed line coincidentally stores the marker value. A practical solution must efficiently handle such collisions, even though such collisions are expected to be extremely rare.  

We propose {\em CRAM}, an efficient hardware-based compression design, that does not require any OS support or changes to the memory module/protocols, and avoids the bandwidth overheads of metadata lookups.  We discuss our evaluation methodology before discussing our solution.


\ignore{
\section{Background and Motivation}

Data compression allows for storing lines in a smaller amount of space and enables retrieving those lines from memory in a fewer number of memory requests or bursts. This enables higher effective main memory bandwidth. We discuss data compression's potential for improving main memory bandwidth, paths for enabling improved bandwidth, mechanisms for locating compressed lines, and considerations in designing a practical and robust main memory compression technique.

\subsection{Potential for Bandwidth Improvement}

To show the potential for compression to enable improved bandwidth, we show the likelihood of lines to compress to half-size and quarter-size in Figure~\ref{fig:compressibility}. 
On average, our workloads show that x\% of lines are compressible to half size, and y\% of lines are compressible to quarter size. If we could read all of the lines with reduced bandwidth, we could save up to x\% of the bandwidth. However, achieving these benefits in a practical manner can be difficult.

\begin{figure}[htb]
	\centering
\centerline{\epsfig{file=FIGS/compressibility.eps, width=\columnwidth}}
	\vspace{-0.08 in}
	\caption{Fraction of compressible lines per workload. x\% of lines compress to half-size.}
	\vspace{-0.13 in}
	\label{fig:compressibility}
\end{figure}

\subsection{How to Compress for Bandwidth}

Prior works have explored various ways to exploit compression to improve memory bandwidth. These works adhere to one of two main approaches. We briefly describe the two approaches below:

\subsubsection{Compact In-place}

One way to improve memory bandwidth is to compress data in-place, and send the data in a fewer number of bursts, similar to MemZip~\cite{shafiee2014memzip}. Figure~\ref{fig:compressibility}(a) shows where 4 quarter-compressible lines A,B,C,D would reside in memory. If we compress them in-place, as in Figure~\ref{fig:compressibility}(b), we could potentially need just 25\% of the bandwidth to read each line.
If we access lines A,B,C,D, we could fetch them in 4 quarter-width accesses and save 75\% of the bandwidth. 

Unfortunately, commodity DDR3 DRAM chips have minimum burst-length requirements of eight. For a standard 64-bit DDR3 bus, this means that we are required to fetch a minimum of 64 bytes (cacheline size) on each access. To achieve these bandwidth savings, we would unfortunately need specialized DRAM with sub-cacheline access granularities. 


\begin{figure}[htb]
	\centering
 \centerline{\epsfig{file=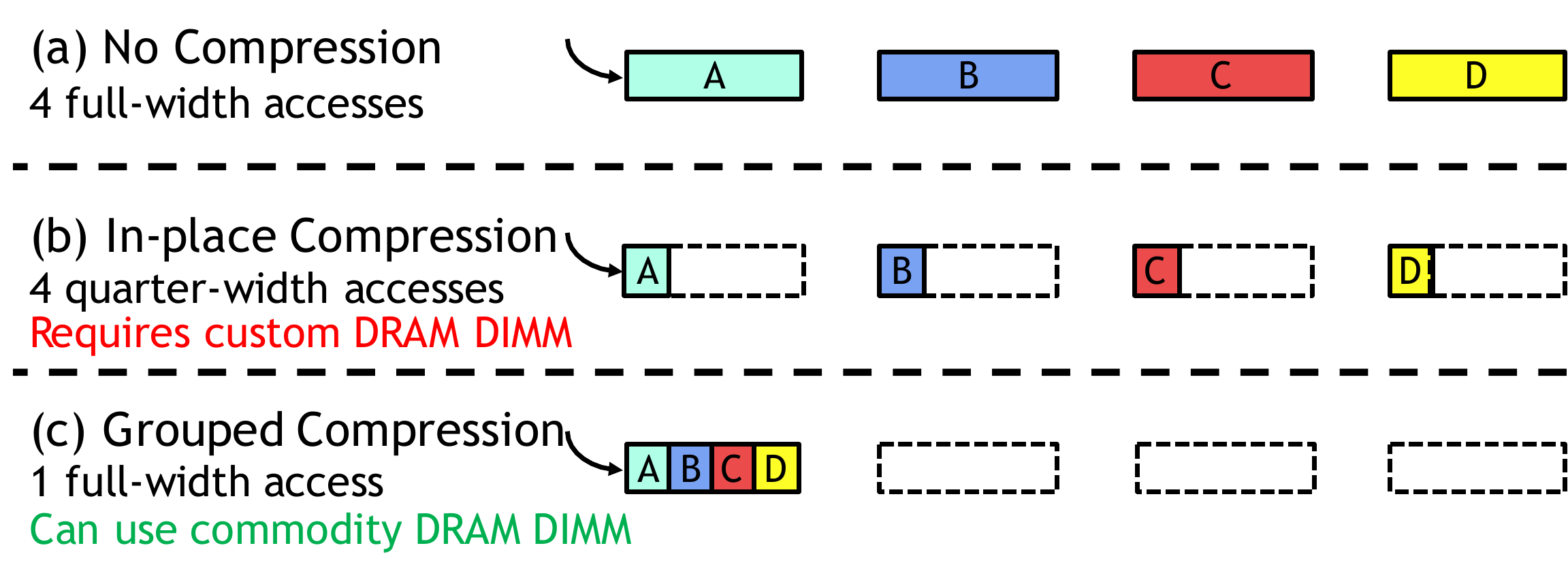, width=\columnwidth}}
	\vspace{-0.08 in}
	\caption{Line placement for (a) uncompressed memory, (b) in-place compression, and (c) ganged compression.}
	\vspace{-0.13 in}
	\label{fig:motivation_bwcompression}
\end{figure}

\subsubsection{Compact Multiple Lines Together}

An alternative method to save bandwidth is to compact multiple adjacent lines together into one location (similar to ~\cite{LCP:MICRO2013}), as shown in Figure~\ref{fig:motivation_bwcompression}(c). If lines A,B,C,D are compressible to quarter size, we could compress them together into one physical location.
On the next access to line A, we would also receive lines B,C,D as well. If we install the lines and they are used, this technique can also end up saving 75\% of the bandwidth, as four requests to A,B,C,D would be serviced with just one request. The benefit of this approach is that it does not require changes to the burst length, so it maintains compatibility with commodity DRAM.
As we aim to have our technique be applicable to commodity DRAM, we take this approach.

Unfortunately, a drawback of this approach is that lines are relocated relative to an uncompressed case--locating the line is made more complex. Such approaches need an effective means to locate and interpret the requested line.

\subsection{How to Locate Compressed Lines}

There are two major approaches to aid in locating compressed lines when they are moved:

\subsubsection{With OS and Page-table Support}

Many prior approaches~\cite{IBM-MXT:HPCA2001,Strenstorm:ISCA2005,LCP:MICRO2013} take the path of extending the page table entries to include information on the compressibility of the page. These approaches are attractive as they allow locating and interpreting lines to be done by the TLB and page table walks, and consequently can be done with little overhead. However, such approaches require OS-changes and other software-support that can limit their applicability. In addition, these approaches can break when applied to systems using large pages. Another limitation is that these approaches inherently require OS-support in order for the system to use the additional capacity offered by compression. We cannot take this approach as we desire a solution that can be implemented without the support of the OS.

\subsubsection{With Metadata Lookup}

An alternate approach is to always look up a separate metadata region in memory to learn 1. how to locate the data, and 2. how to interpret the data received. Memzip~\cite{shafiee2014memzip} probes and updates the metadata to know how many bursts it will need to read a compressed line. LCP~\cite{LCP:MICRO2013} also does a metadata lookup to help interpret what line it has received, as well as to locate incompressible lines. And both approaches use a metadata cache to reduce the amount of accesses to the metadata region. These approaches work well when the metadata cache has a high hit-rate, due to either high spatial locality or small workload footprint. However, these approaches break when scaled to much larger workloads with low spatial locality. At the worst-case, these approaches may need a separate metadata access for every data access, constituting a potential bandwidth overhead of 100\%. If we are to build a robust main memory compression scheme, we will need to solve this overhead.

Unfortunately, neither approach is currently viable for making a robust OS-transparent compressed main memory.
    
\subsection{Prior Work and Our Approach}

We compare with prior works that target improving main memory bandwidth with compression in Table~\ref{tab:motivation_table}.

\begin{table}[htb]
    \begin{center}
    \caption{Prior approaches for improving memory bandwidth}
    \vspace{-0.08 in}
\centerline{\epsfig{file=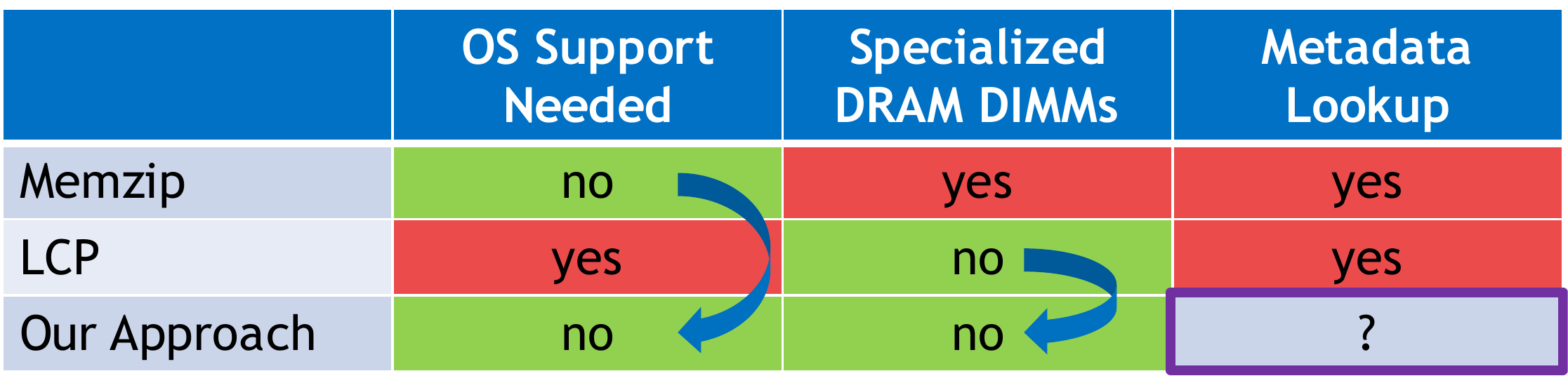, width=\columnwidth}}
	\vspace{-0.13 in}
	\label{tab:motivation_table}
    \end{center}
\end{table}

    Memzip~\cite{shafiee2014memzip} compresses data in-place and uses smaller burst-lengths to try to achieve bandwidth benefits. Memzip explicitly looks up metadata to learn compressibility and burst-length, and it sacrifices capacity benefits to maintain OS-transparency. However, it needs non-commodity DRAM to achieve its bandwidth benefits. We take its approach for OS transparency--we sacrifice capacity benefits in order to retain OS transparency.

	LCP uses packs multiple contiguous lines together to get bandwidth benefits on commodity DRAM. However, it requires OS support and metadata lookup to interpret the lines. We take a similar approach and pack together multiple lines so that our scheme can work on commodity DRAM.
    
	However, we must solve the key flaw of metadata lookup if we are to maintain robustness when workload footprints scale and metadata cache hit-rates become poor.
    
\subsection{Insight: Store Metadata in Space Saved by Compression}

A key difficulty in compressed main memory is that it has no space for tags or metadata to help interpret the 64 bytes that is read. If we could find some way to store information on how to interpret the line that was read, we could avoid the need to do a separate metadata lookup. Luckily, we have an insight that as compression saves space and often has un-used space, we can in-line compressibility information in this un-used space, as shown in Figure~\ref{fig:motivation_metadata}.
We could, for example, store a special 4 byte marker in every compressed location. On a read to a location that has a matching 4 byte marker, we could directly interpret the location as a compressed location and obviate the need for a separate metadata lookup. Unfortunately, we would still be wrong on average one every 4 billion accesses in the case an uncompressed line happens to store a value matching this 4 byte compressibility marker.
Throughout this paper, we handle the corner cases and enable a solution that completely removes the cost of metadata lookup, resulting in a main memory compression technique that improves bandwidth, does not degrade performance, and can be done with commodity DRAM and without OS support.

    \begin{figure}[htb]
	\centering
\centerline{\epsfig{file=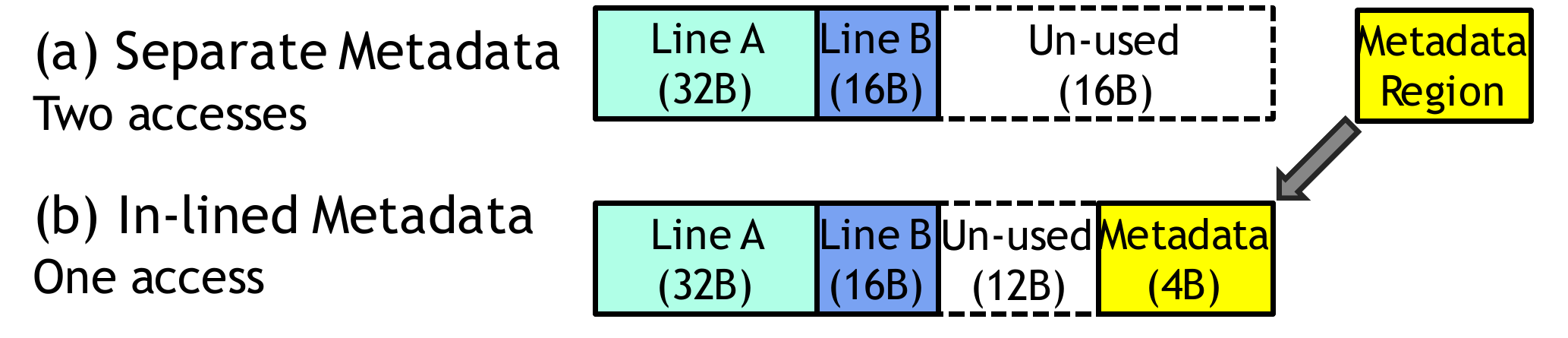, width=\columnwidth}}
	\vspace{-0.08 in}
	\caption{In-lining metadata into un-used space of compressed lines reduces need for metadata lookup.}
	\vspace{-0.13 in}
	\label{fig:motivation_metadata}
\end{figure}

We could, for example, store a special 4 byte marker in every compressed location. On a read to a location that has a matching 4 byte marker, we could directly interpret the location as a compressed location and obviate the need for a separate metadata lookup.

\textbf{Why it works:} If we read a line that has this 4-byte marker, there is a high likelihood that this line was a previously compressed line. But, storing this 4-byte marker can come at the cost of compression ratio. Fortunately, we find that most lines do not fully use the 64-byte space and have enough space to store this 4-byte marker. Figure~\ref{fig:motivation_insight} shows the probability of a pair or quad of adjacent lines compressing to $\leq$64B and $\leq$60B. As the probability of compressing pairs of lines to $\leq$64B and $\leq$60B are 38\% and 36\%, respectively, we find that reserving space for this marker does not substantially impact likelihood of compressing lines together and thus does not significantly impact compression ratio.


\textbf{Caveat:} Unfortunately, this marker-based approach could still be wrong on average one every 4 billion accesses in the case an uncompressed line happens to store a value matching this 4 byte compressibility marker.
Throughout this paper, we handle the corner cases (by inverting colliding lines) and enable a solution that completely removes the cost of metadata lookup, resulting in a main memory compression technique that improves bandwidth, does not degrade performance, and can be done with commodity DRAM and without OS support.

}

  
\ignore{
\section{Background and Motivation}
\subsection{Exploiting compression for bandwidth}

The basic idea behind using compression for bandwidth is that data compaction allows for the retrieval of data from the main memory in fewer memory requests which translates to a greater effective main memory bandwidth.

The analysis of workloads from the spec2006 suite shows that a large number of workloads are compression friendly. Fig 3a,b shows a breakdown of the compressibility of cachelines from the SPEC2006 \& SPEC2017 benchmark suite. We categorize the cachelines as being incompressible, 2-way or 4-way compressible with their neighboring cachelines. The large percentage of compressible cachelines in these workloads shows the potential for bandwidth savings that can be obtained with compression.

Prior works have explored various ways to exploit compression for memory bandwidth. These works adhere to one of two main design points. We briefly describe the two approaches below:
\subsubsection{In-place Compression}
This scheme performs data compression while keeping cachelines in their original physical location. Bandwidth improvement is achieved by issuing fewer memory bursts to access the data. Lets explain this scheme with the help of an example. Consider four adjacent cachelines A, B, C, D located in memory as shown in Figure~\ref{fig:motivation_bwcompression}(a). In an uncompressed memory system, we require 4 full-width memory accesses to fetch these lines. Now, let us assume that these cachelines are compressible to one-fourth the original size. With in-place compression, the location of these cachelines remain unchanged  (Figure~\ref{fig:motivation_bwcompression}(b)) \& each of these lines can now be fetched with quarter-width memory bursts. Thus, four compressed lines can be retrieved at the cost of one full-width memory request.

In-place compression avoids the problem of having variability in the location of the cacheline. The main drawback however is that it requires the memory system to support fractional burst length. Due to this requirement, in-place compression breaks compatibility with current dram memory systems. Memzip is an example of the in-place compression scheme. It uses a rank-subsetting based memory to support sub-cacheline burst lengths. This however would need major changes to the dram memory, controller and the interface and would break compatibility with current dram main memory systems.

\begin{figure}[htb]
	\centering
	\centerline{\epsfig{file=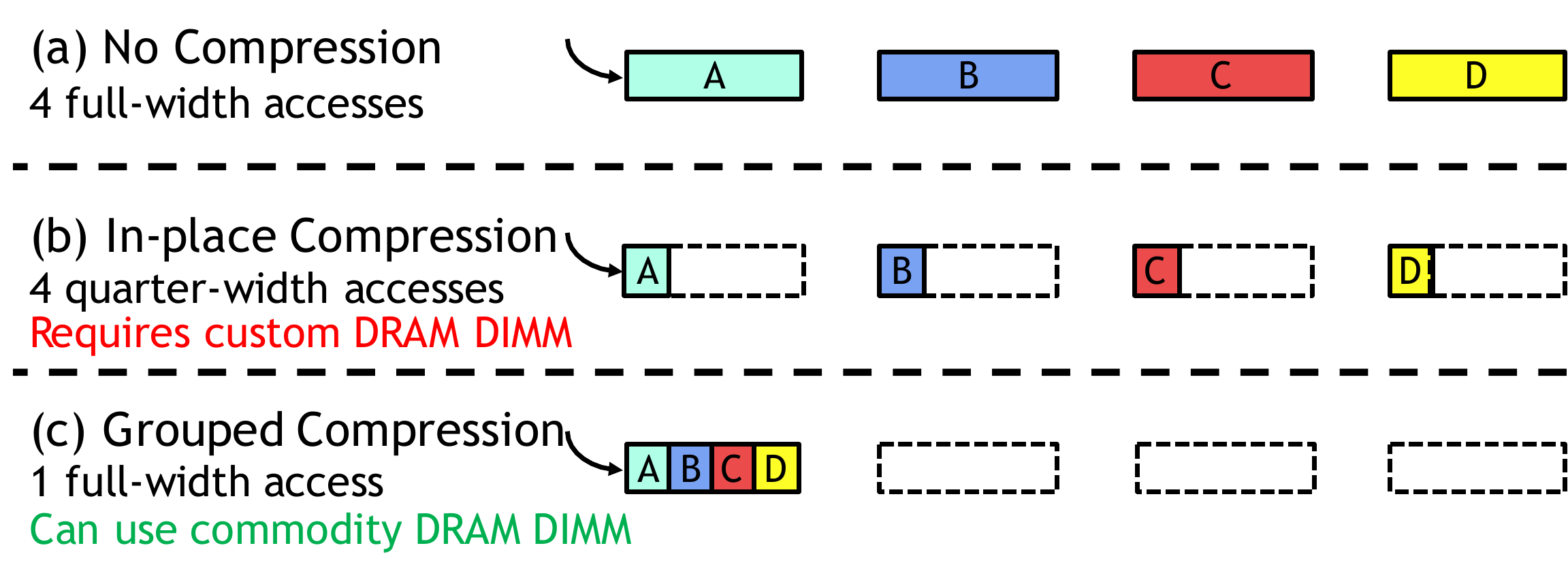, width=\columnwidth}}
	\caption{a. b. c.}
	
	\label{fig:motivation_bwcompression}
\end{figure}

\subsubsection{Ganged Compression}
In a ganged compression scheme, we group together adjacent cachelines and co-locate them in one physical location so that multiple cachelines can be retrieved with one request. Continuing with our example from before, in case of ganged-compression, we group cachelines A,B,C,D so that they are resident in the same physical address as shown in Figure~\ref{fig:motivation_bwcompression}(c). A memory request to cacheline A would be able to prefetch cachelines B,C,D at no extra bandwidth cost. So, this in effect would translate to 4X the original bandwidth if the prefetches are useful. In contrast to the in-place compression scheme, this does not require changes to the burst length, so it maintains compatibility with commodity dram memory systems. LCP is an example of a prior work that is based on the principles outlined in this scheme. Each cacheline in a page is expected to be compressible to a specific size. LCP allocates pages of different sizes based on the compressibility of cachlines. On sending a request to a compressed page, additional cachelines can be prefetched at no extra bandwidth cost. The benefit of using this scheme is that it does not require changes to the dram interface as it works with standard burst sizes. Due to this reason we use the ganged-compression scheme in our work to maintain compatibility with commodity dram memory.

The main challenge with implementing a ganged compression scheme is the variable location of cachelines. It is clear from our example in Fig ~\ref{fig:motivation_bwcompression}(c) that we have to shift cachelines B,C,D away from their original physical addresses. If in case these cachelines were not compressible, they would be located in their original physical address. So the location of cachelines changes based on their compressibility. Schemes using ganged-compression thus need a way to determine compressibility before accessing the cacheline.  LCP handles the problem of locating  cacheline by seeking the support of the operating system to handle  allocation and management of variable sized pages to support compression.

}



  \ignore{
  Flow:
  1. Compression for bandwidth - graph - workloads are compressible
  2. Two ways of exploiting compression for bandwidth a. fetch multiple lines b. send fewer burst to fetch only the necessary amount of data
  (mention adv, disadv of each)
  Prior works adhere to one of these two approaches 
  3. Describe Memzip and LCP

  }

\ignore{
\subsection{Prior works on memory compression}
There are multiple works which explore main memory compression for bandwidth. We would like to discuss two main works which most closely relate to our paper.
\subsubsection{Memzip}
Memzip enables memory compression for bandwidth by using the in-place compression approach. A rank-subsetting based dram memory is used as the baseline to implement memory compression. This involves changing the way data is organized in a typical dram system. Instead of spreading a cacheline across multiple ranks, rank-subsetting places the entire cacheline in a single rank. By allowing for memory access at a fine grained granuality, requests can be issued to fetch only a fraction of the original uncompressed size of the cacheline. With compression, fewer requests can be issued to retrieve the compressed cacheline which improves the effective main memory bandwidth.

Rank subsetting is a non trivial change to the main memory. It would involve major modifications to the existing dram memory architecture and would break compatibility with existing dram interface. Memzip also suffers from having to do metadata lookups. Although the location of cachelines are known, the number of bursts that are needed to retrieve the cacheline needs to be determined. The number of bursts needed depends on the compressibility of the line and this information needs to be stored in a metadata table. Lookups to this table causes additional bandwidth overheads which degrades performance. 
\subsubsection{Linearly Compressed Pages (LCP)}
LCP achieves higher memory bandwidth \& capacity by the principles outlined in the ganged compression scheme. The bandwidth benefits of compression are exploited by bandwidth-free prefetches of adjacent cachelines as described before. 
Each cacheline in a page is expected to be compressible to a specific size. LCP allocates pages of different sizes based on the compressibility of cachlines. In case the line is incompressible, it get placed in a special exception region of the page and its location is tracked using metadata bits.

LCP requires the support of the operating system to allocate and track pages of different sizes in memory. It is also reliant on a metadata cache to locate lines in the exception region. Each memory request thus needs to check the metadata associated with the page first. A metadata miss triggers multiple requests which incur additional bandwidth and latency overheads. This ends up hurting performance significantly for workloads having memory accesses with low spatial locality.
}

\ignore{
Vinson flow

Potential for Compressing for bandwidth
	Fraction 2. 4. 8-way compressible. can reduce x\% of bandwidth if all brought in at once. Effective bw of 1.7x if all lines are useful.
    
    FIGURE COMPRESSIBILITY

How to get bandwidth:
	1. compress in-place. save bursts. but, need sub-cacheline burst lengths~\cite{shafiee2014memzip}. (bad) need specialized dram.
    
    2. Alternatively, spatially compress multiple lines together. same minimum burst length, but get additional lines without bandwidth cost~\cite{LCP}. protocol-compliant. If we are to avoid needing specialized DRAM, we should go this path.
    
    Problem: multiple possible locations. need effective mechanisms to locate line.
    
How to locate moved lines:
	1. OS support / page-table (mxt, stenstrom, lcp). in-lines compressibility information in page table entries. needs os-support, not as easily applicable.
    
    2. Additional metadata (lcp, memzip). Look up separate metadata region, before accessing. Metadata cache works for workloads with spatial locality. But, as workloads scale up, metadata cache will have low hit-rate, and constantly looking up this metadata can degrade performance.
    
    No good solution exists.
    
Pior works and our approach
	FIGURE COMPARISON

    Memzip explicitly looks up metadata and sacrifices capacity benefits to maintain OS-transparency. However, it needs non-commodity DRAM to get bandwidth benefits. We similarly sacrifice capacity benefits for OS transparency.
    LCP uses packs multiple contiguous lines together to get bandwidth benefits on commodity DRAM. However, it requires OS support and metadata lookup to interpret the lines. We similarly pack together multiple lines to achieve applicability to commodity DRAM.
	However, we must solve for the metadata lookup in order to retain robustness when workload footprints are scaling and metadata cache hit-rates are poor.
    
Insight: Compressed lines have space for metadata. Can remove separate metadata lookup.
	FIGURE |0-31|32-48|unused|60-63 metadata
	For example, store 4B code saying it is a compressed line. Only 1 in 2^32 incorrect classification, which we will handle. Access to the line also gets information on how to interpret line. avoid separate metadata lookup.
}

\ignore{

\subsection{Problem: BW-intensive Metadata Lookup}

\ignore{
Problem: compression moves lines. Need way to interpret lines that are read (compressibility and location). No additional space per line to help interpret. Traditionally, keep track of exact information compressibility and location in metadata region~\cite{}. For workloads with poor spatial locality, this metadata lookup can consume additional 50-80\% bandwidth (incompressible workloads with poor spatial locality, such as xz or cactu).
}

A conventional approach to handle interpreting lines is to maintain compression and location information in a separate metadata region in memory to remember 1. how to locate the data, and 2. how to interpret the data received. For example, Memzip~\cite{shafiee2014memzip} probes and updates the metadata to know how many bursts it will need to read a compressed line. Another example, LCP~\cite{LCP:MICRO2013}, also does a metadata lookup to help interpret what line it has received, as well as to locate incompressible lines. And both approaches use a metadata cache to reduce the amount of accesses to the metadata region. These approaches work well when the metadata cache has a high hit-rate, due to either high spatial locality or small workload footprint. However, these approaches break when scaled to much larger workloads with low spatial locality. At the worst-case, these approaches may need a separate metadata access for every data access, constituting a potential bandwidth overhead of 50-80\% (e.g., in \textit{xz} and \textit{cactu}). If we are to build a robust main memory compression scheme, we will need to solve this overhead.

}


%% file: methodology.tex
\section{Methodology}

\subsection{Framework and Configuration}
\label{subsection:conf}

We use USIMM~\cite{USIMM}, an x86 simulator with detailed memory
system model.  
Table~\ref{table:config} shows the configuration used in our study.
We assume a three-level cache hierarchy (L1, L2, L3 being on-chip SRAM
caches). All caches use line-size of 64 bytes. The DRAM model is based on DDR4. 

We model a virtual memory system to perform virtual to physical
address translations, and this ensures that the memory accesses of different cores do not map to the same physical page.  Note that, other than the virtual memory translation, the OS is not extended to provide any support to enable the compressed memory.

For compression, we use a hybrid compression scheme where we use FPC and BDI and compress with the one that gives better compression. Information about the compression algorithm used and the compression-specific metadata (e.g. base for BDI) are stored within the compressed line, and are counted towards determining the size of the compressed line.

\begin{table}[ht]
  \begin{center}
      \caption{System Configuration}

\renewcommand{\arraystretch}{.80}
\setlength{\extrarowheight}{2pt}{
      \begin{tabular}{|l|l|}  \hline 

Processors & 8 cores; 3.2GHz, 4-wide OoO \\
Last-Level Cache & 8MB, 16-way \\ 
Compression Algorithm & FPC + BDI \\ \hline

\multicolumn{2}{|c|}{\bf Main Memory } \\ \hline

Capacity               &    16GB         \\
Bus Frequency          &     800MHz (DDR 1.6GHz) \\
Configuration       &  2 channel, 2x rank, 64-bit bus \\
tCAS-tRCD-tRP-tRAS     &      11-11-11-39 ns \\ \hline

\ignore{
\multicolumn{2}{|c|}{\bf Main Memory (PCM) } \\ \hline

Capacity               &    64GB        \\
Bus Frequency          &     1000MHz (DDR 2GHz) \\
Configuration       &  2 channel, 64-bit bus \\
Aggregate Bandwidth              &      32 GB/s \\ 
tCAS-tRCD-tRP    &      13-128-8 ns \\
tRAS-tWR     &          143-160 ns \\ \hline
}
      \end{tabular}
}
      \label{table:config}
\vspace{-0.15 in}

      \end{center}
\end{table}

\newpage
\subsection{Workloads}
\label{subsection:workloads}

We use a representative slice of 1-billion instructions selected by
PinPoints~\cite{pinpoint}, from benchmarks suites that include SPEC
2006~\cite{SPEC2006}, SPEC 2017~\cite{spec17}, and GAP~\cite{GAP}. 
We evaluate all SPEC 2006 and SPEC 2017 workloads, and mark '06 or '17 to denote the version when the workload is common to both.
We additionally run GAP suite, which is graph analytics 
with real data sets (\texttt{twitter}, \texttt{web sk-2005})~\cite{Davis:matrix2011}. We show detailed evaluation of the workloads with at least five misses per thousand instructions (MPKI). The evaluations execute benchmarks in rate mode, where all
eight cores execute the same benchmark.
Table~\ref{table:benchmarks}
shows L3 miss rates and memory footprints of the workloads we have evaluated in detail.
In addition to these workloads, we also include 6 mixed workloads that are formed by randomly mixing the SPEC workloads.


\begin {table}[ht]

\caption{Workload Characteristics}
\begin{center}
\renewcommand{\arraystretch}{.92}
\setlength{\extrarowheight}{2pt}{
        \begin{tabular}{|c|c|c|c|} \hline

Suite & Workload &  L3 MPKI  & Footprint \\ \hline \hline

\multirow{15}{*}{SPEC} & fotonik & 26.2  & 6.8 GB  \\ \cline{2-4}
         & lbm17    & 25.5  & 3.4 GB  \\ \cline{2-4}
         & soplex  & 23.3  & 2.1 GB  \\ \cline{2-4}
         & libq    & 23.1  & 418 MB  \\ \cline{2-4}
         & mcf17     & 22.8  & 4.4 GB  \\ \cline{2-4}
         & milc    & 21.9  & 3.1 GB  \\ \cline{2-4}
         & Gems    & 17.2  & 5.8 GB  \\ \cline{2-4}
         & parest  & 16.4  & 465 MB  \\ \cline{2-4}
         & sphinx  & 11.9  & 223 MB  \\ \cline{2-4}
         & leslie  & 11.9  & 861 MB  \\ \cline{2-4}
         & cactu17   & 10.6  & 2.1 GB  \\ \cline{2-4}
         & omnet17   & 8.6   & 1.9 GB  \\ \cline{2-4}
         & gcc06     & 5.8   & 205 MB  \\ \cline{2-4}
         & xz      & 5.7   & 943 MB  \\ \cline{2-4}
         & wrf17     & 5.2   & 798 MB  \\ \hline
\multirow{6}{*}{GAP}      & bc twi  & 66.6  & 9.2 GB  \\ \cline{2-4}
         & bc web  & 7.4   & 10.0 GB \\ \cline{2-4}
         & cc twi  & 101.8 & 6.0 GB  \\ \cline{2-4}
         & cc web  & 8.1   & 5.3 GB  \\ \cline{2-4}
         & pr twi  & 144.8 & 8.3 GB  \\ \cline{2-4}
         & pr web  & 13.1  & 8.2 GB  \\ \hline

\ignore{
\multirow{7}{*}{SPEC2006} & soplex  & 23.3  & 2.1 GB  \\ \cline{2-4}
         & libq    & 23.1  & 418 MB  \\ \cline{2-4}
         & milc    & 21.9  & 3.1 GB  \\ \cline{2-4}
         & Gems    & 17.2  & 5.8 GB  \\ \cline{2-4}
         & sphinx  & 11.9  & 223 MB  \\ \cline{2-4}
         & leslie  & 11.9  & 861 MB  \\ \cline{2-4}
         & gcc     & 5.8   & 205 MB  \\ \hline
\multirow{8}{*}{SPEC2017} & fotonik & 26.2  & 6.8 GB  \\ \cline{2-4}
         & lbm     & 25.5  & 3.4 GB  \\ \cline{2-4}
         & mcf     & 22.8  & 4.4 GB  \\ \cline{2-4}
         & parest  & 16.4  & 465 MB  \\ \cline{2-4}
         & cactu   & 10.6  & 2.1 GB  \\ \cline{2-4}
         & omnet   & 8.6   & 1.9 GB  \\ \cline{2-4}
         & xz      & 5.7   & 943 MB  \\ \cline{2-4}
         & wrf     & 5.2   & 798 MB  \\ \hline
\multirow{6}{*}{GAP}      & bc twi  & 66.6  & 9.2 GB  \\ \cline{2-4}
         & bc web  & 7.4   & 10.0 GB \\ \cline{2-4}
         & cc twi  & 101.8 & 6.0 GB  \\ \cline{2-4}
         & cc web  & 8.1   & 5.3 GB  \\ \cline{2-4}
         & pr twi  & 144.8 & 8.3 GB  \\ \cline{2-4}
         & pr web  & 13.1  & 8.2 GB  \\ \hline
         }

        \end{tabular}
}
      \label{table:benchmarks}
    \end{center}
\end{table}

We perform timing simulation until each benchmark in a workload
executes at least 1 billion instructions.  We use weighted speedup
to measure aggregate performance of the workload normalized to the
baseline and report geometric mean for the average speedup across all
the 27 workloads (7 SPEC2006, 8 SPEC2017, 6 GAP, 6 MIX). For other workloads that
are not memory bound, we present full results of all 64 benchmarks evaluated (29 SPEC2006, 23 SPEC2017, 6 GAP, 6 MIX) in Section~\ref{ssec:all_workloads}.

\clearpage




%% file: organization.tex
\ignore{
FLOW:
- Organization
- Placement of lines x2 and x4
- Operation (what happens on read and write)

}

\section{CRAM: Basic Design}

Our proposed design, CRAM, tries to obtain bandwidth benefits using memory compression without requiring OS support, without changes to bus protocol, and while maintaining the existing organization for the memory modules.  In this section, we provide an overview of the basic CRAM design, and discuss the shortcomings of maintaining and retrieving metadata associated with compression.

\subsection{Organization and Operation of CRAM}

\begin{figure}[htb]
\vspace{-0.13 in}
	\centering
 \centerline{\epsfig{file=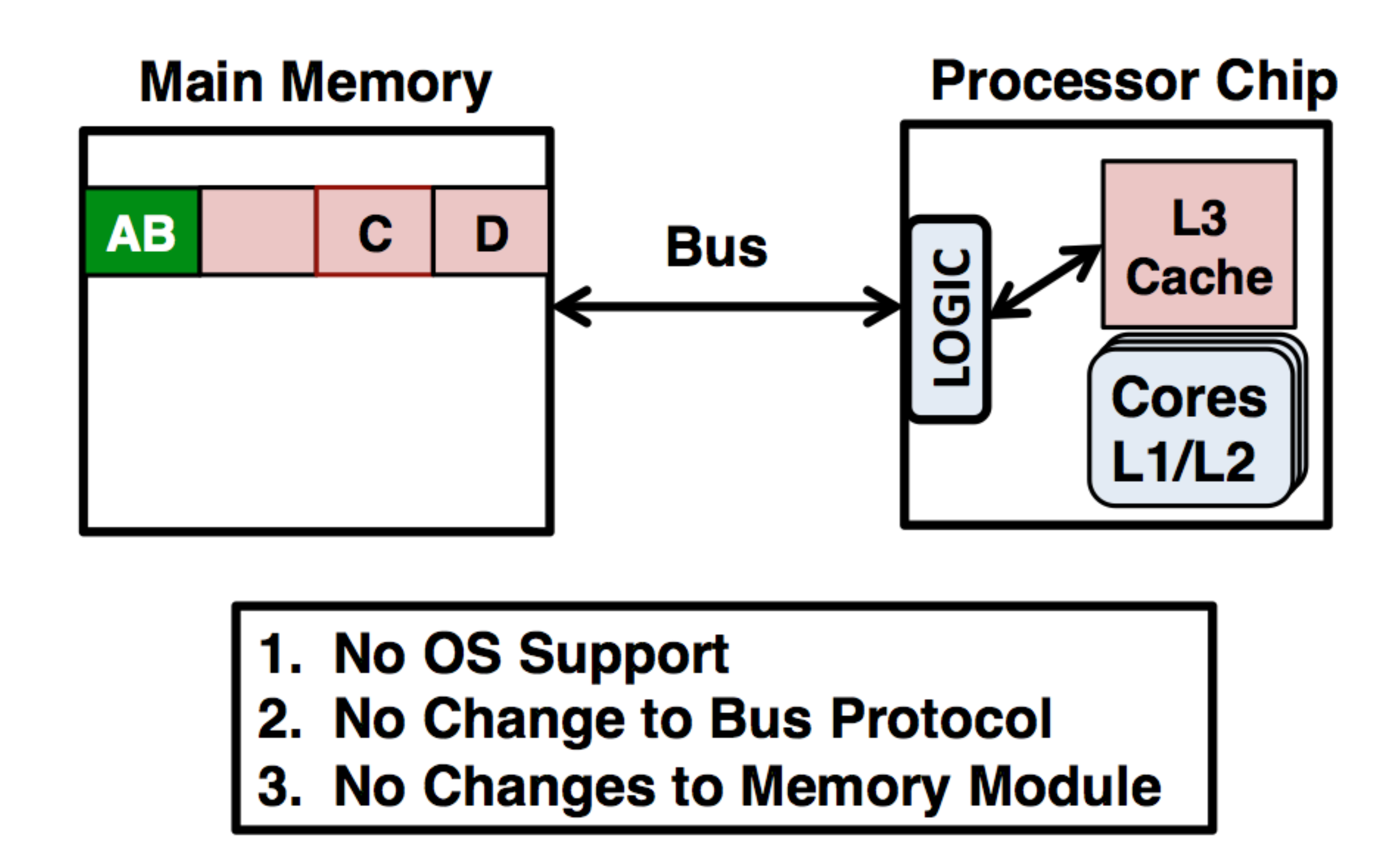, width= 3 in}}
 \vspace{-0.1 in}
	\caption{An Overview of the CRAM Design.}
	\label{fig:cramarch}
\end{figure}

Figure~\ref{fig:cramarch} shows an overview of CRAM.  The main memory can store compressed data, and the job of compression and decompression is performed by the logic on the memory controller on the processor chip.  The L3 cache is assumed to store data in uncompressed form.  The bus connecting the memory controller and the memory modules use the existing JEDEC protocol and transfer 64 bytes on each access. If the lines are compressible, then a single access can provide multiple neighboring lines, and increase effective bandwidth.

\noindent\textbf{Restricted Data Mapping:}   Without loss of generality, CRAM supports up to 4-to-1 compression, which means up to four compressed lines can be resident in one memory location.  If the compressibility is not high enough to store 4 lines in one location, then the design tries 2-to-1 compression, where two neighboring lines are placed in one location. If the lines are uncompressed they retain their original location. If we organize the data layout appropriately, we can reduce the amount of uncertainty (i.e., number of possible positions) in locating lines. CRAM restricts the location of the lines in a group of 4 lines to help in locating the lines easily.

\begin{figure}[htb]
	\centering
 \centerline{\epsfig{file=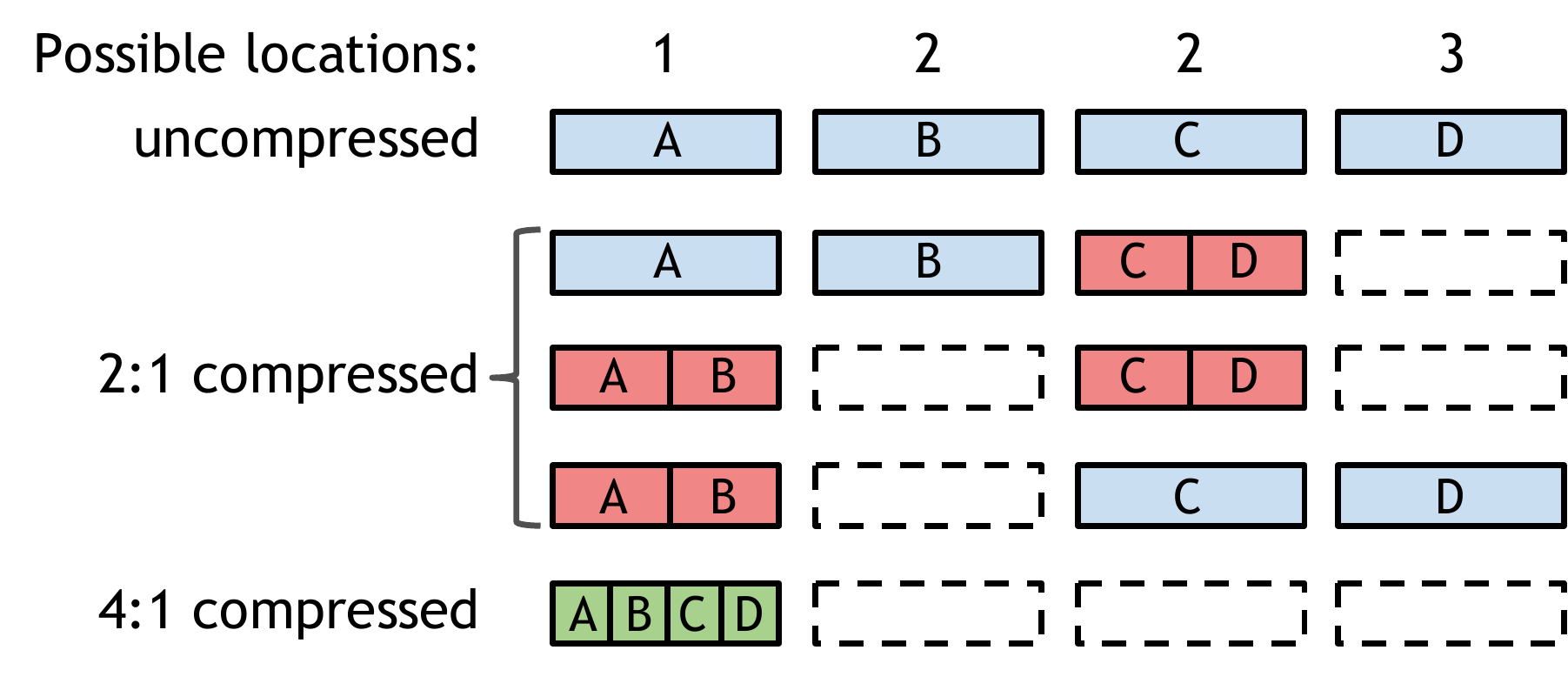, width=3in}}
	\caption{CRAM relocates and packs adjacent lines. Restricting placement reduces \ignore{total }number of valid positions.}
	\label{fig:indexing}
\end{figure} 

Figure~\ref{fig:indexing} shows the five different line permutations for a group of 4 lines under CRAM, based on whether the lines undergo 2-to-1 compression, 4-to-1 compression, or no compression. Thus, line A (lines with line-address ending in "00") is always resident in the same location, whereas line B (lines with address ending in "01") can be in the original location at B (if B is uncompressed) or at A (if B is compressed). Note that, on average there are only two locations for each line in the group.  An access to line A can provide location information for all four lines in the group if the line is 4-to-1 compressed, or for line B otherwise.  Thus, a sequential access across the memory would obtain the first line in the group always from the original location, and this line can provide location information for the subsequent lines in the group.


\vspace{0.05 in}
\noindent\textbf{Write Operation:} When a cacheline gets evicted from LLC, the memory controller checks if (a) the neighboring cachelines are present in the LLC, and if (b) the group of 2 or 4 cachelines can be compressed to the size of a single uncompressed cacheline (64 Bytes). If the group of cachelines can be compressed to a single block, the memory controller compacts them together and issues a write containing the 2 or 4 compressed lines to one physical location. Note that for a compressed memory, the controller can have the flexibility to compress and write back \textit{clean} lines as well, otherwise the benefits of compression will become restricted only to dirty lines. Our default policy compacts and writes back clean lines if they are compressible, in the hope that this bandwidth cost will be amortized by future re-use.

\vspace{0.05 in}
\noindent\textbf{Read Operation:} On a read, the controller needs to determine the compressibility and the location of the line, as the line can get relocated based on compressibility.  Conventional designs rely on metadata in memory to provide the {\em Compression Status Information (CSI)} of each line.  In our case, this CSI-metadata would be a 3-bit entity for a group of 4 lines (to indicate one of 5 possible states for the group, based on Figure~\ref{fig:indexing}).  If the CSI metadata is available, the read can determine the location of the line, access the memory, decompress all the line(s) if needed, and store all the retrieved lines in the L3 cache.

\begin{figure}[htb]
	\centering
    \vspace{-.17in}
\centerline{\epsfig{file=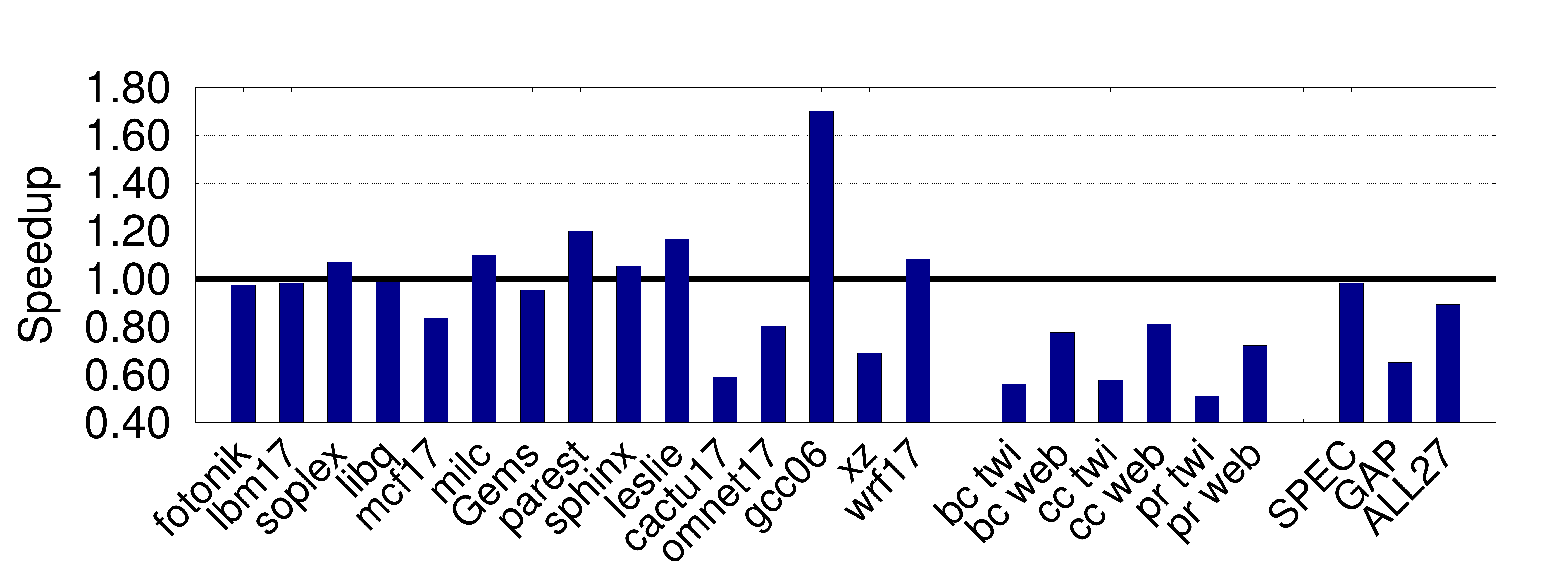, width=\columnwidth}}
	\vspace{-0.11 in}
	\caption{Speedup of CRAM with explicit metadata (+ metadata cache) compared to uncompressed memory.}
	\vspace{-0.23 in}
	\label{fig:naive_performance}
\end{figure}

\begin{figure*}[htb]
    \vspace{-.25in}
	\centering
\centerline{\epsfig{file=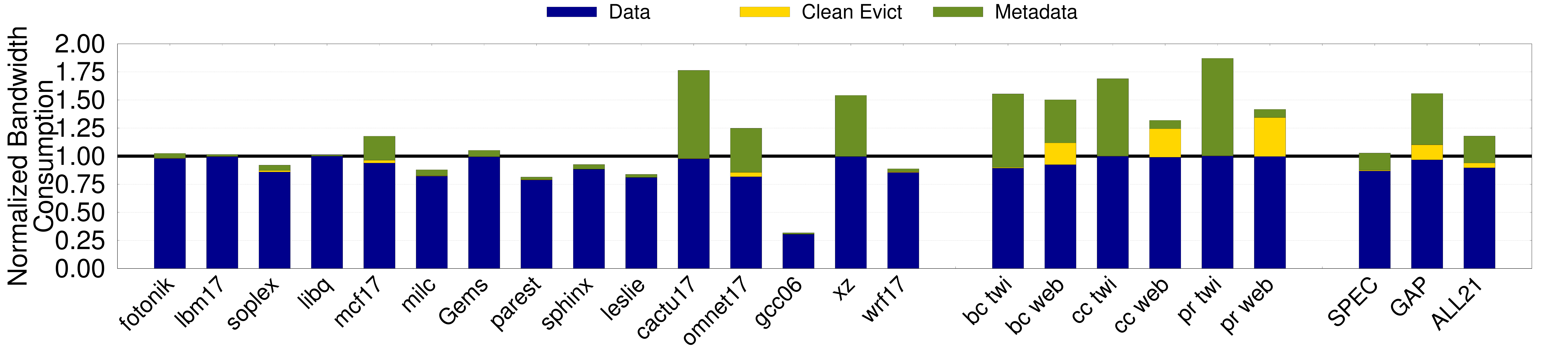, height = 1.6 in}}
	\vspace{-0.15 in}
	\caption{Bandwidth consumption for data, compressed writebacks, and metadata for CRAM with explicit metadata, normalized to uncompressed memory. Metadata accesses constitute significant bandwidth overhead.}
	\vspace{-0.18 in}
	\label{fig:naive_bandwidth}
\end{figure*}

\subsection{The Problem With Explicit Metadata}

We can design CRAM with explicit metadata, where the metadata specifies the compression status of the line. Given that the size of the metadata is 3-bits per group of four lines, we need on average 0.75 bits per line.  For our 16GB memory, containing 1 billion lines, the total size of this metadata would be 24 megabytes, much larger than the capacity of on-chip structures. We can keep the metadata in memory and cache it in an on-chip metadata cache, as done in prior works~\cite{LCP:MICRO2013,shafiee2014memzip}. For workloads with good spatial locality or small memory footprint, most metadata requests and updates will be serviced by the cache and such a design would work well.  Unfortunately, having an explicit metadata requires accessing memory on a miss in the metadata cache.  Figure~\ref{fig:naive_performance} shows the performance of CRAM with explicit metadata with a 32KB metadata cache. On average, this scheme shows 10\% slowdown relative to an uncompressed memory, because of the metadata accesses.  


Figure~\ref{fig:naive_bandwidth} shows the breakdown of the bandwidth consumed by the CRAM design, normalized to the uncompressed memory.  In general, compression is effective at reducing the number of requests for data. However, depending on the workload, the metadata cache can have poor hit-rate, and require frequent access to obtain the metadata. These extra metadata accesses can constitute a significant bandwidth overhead. For example, \textit{xz} needs over 50\% extra bandwidth just to fetch the metadata. Thus, schemes that require a separate metadata lookup can end up degrading performance relative to uncompressed memory. We develop a solution that eliminates metadata lookups.


\vspace{-.1in}
\section{CRAM: Optimized Design}

To avoid the bandwidth overheads of the metadata access, CRAM decouples the information provided by the metadata into two parts: (a) \textit{determining the compression status}, and (b) \textit{determining the location} of the cachelines, and solves each of them separately. The first solution tackles the problem of interpreting accessed lines with implicit-metadata using marker values. The second component handles the problem of the locating lines with a line-location predictor. 


\vspace{-.1in}
\subsection{Implicit-Metadata: No Metadata Lookups}

We exploit the insight that a pair of compressed lines does not always use all the available 64 bytes.  Our analysis (presented in Figure~\ref{fig:motivation_insight}) shows that the probability that a pair of lines is compressible within 64 bytes but not within 60 bytes is quite small (close to 2\%).  We exploit this leftover space in compressed lines to specify the compression status of the line, using a predefined special value, which we call as the {\em marker}. 

\begin{figure}[htb]
	\centering
    \vspace{-0.17 in}
 \centerline{\epsfig{file=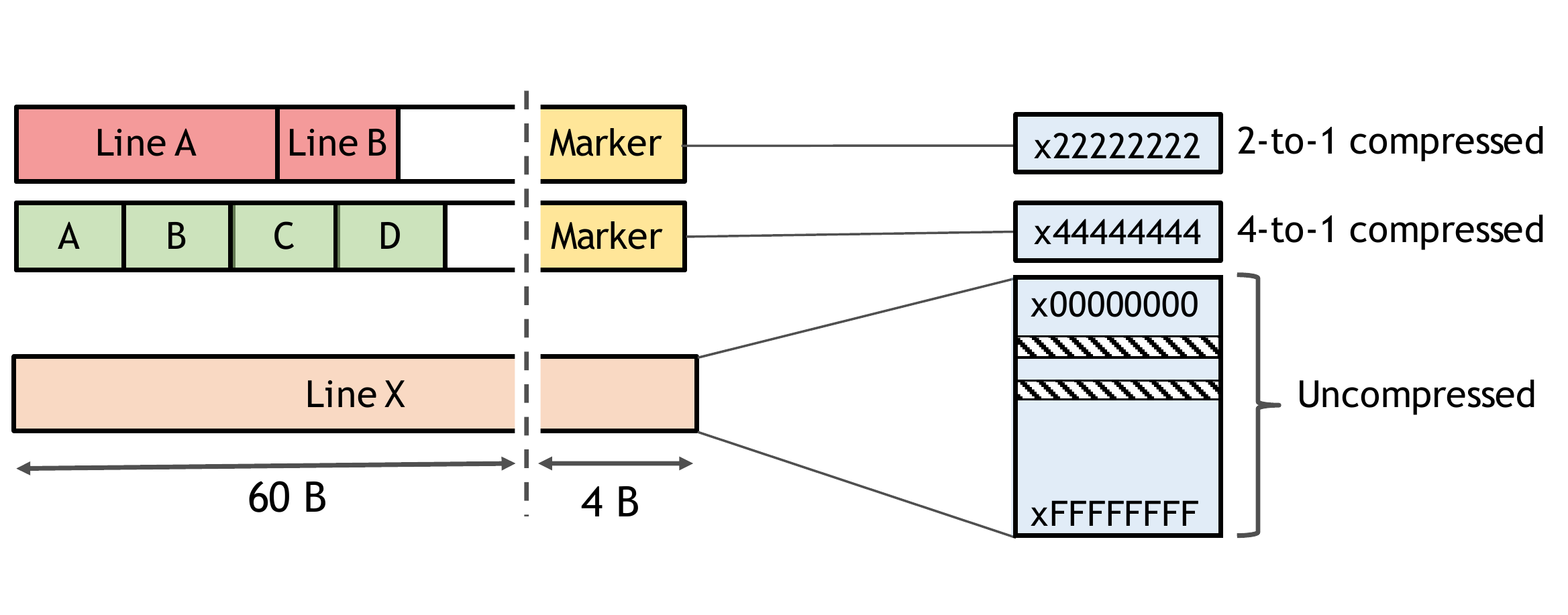, width=\columnwidth}}
	\vspace{-0.20 in}
	\caption{Implicit metadata using markers: Compressed lines always contain a marker in the last four bytes.}
	\vspace{-0.07 in}
	\label{fig:marker}
\end{figure}

Figure~\ref{fig:marker} shows the implicit-metadata design using markers, for lines that are compressible with 2-to-1 compression, 4-to-1 compression, or no compression. If the line contains two compressed lines (e.g. A and B both reside in A), then the line is required to contain the marker corresponding to  2-to-1 compression (x22222222 in our example) in the last four bytes. Similarly, if the line contains four compressed lines (A, B, C, and D, all reside in A), then the line is required to contain the marker corresponding to 4-to-1 compression (x44444444 in our example). Marker reduces the available space for compressed lines to 60 bytes. If the compressed lines require cannot fit within 60 bytes, then it is stored in uncompressed form.  

An incompressible line is stored in its original form, without any space reserved for the marker. The probability that the uncompressed line coincidentally matches with the 32-bit marker is quite small (less than 1 in a billion). Our solution handles such extremely rare cases of collision with marker values by storing such lines in an inverted form.  This ensures that the only lines in memory that contain the marker value (in the last four bytes) are the compressed lines.

\noindent{\bf Determining Compression Status with Markers:}  When a line is retrieved, the memory controller scans the last four bytes for a match with the markers.  If there is a match with either the 2-to-1 marker or the 4-to-1 marker, we know that the line contains compressed data for either two lines, or four lines, respectively. If there is no match, the line is deemed to store uncompressed data.  Thus, with a single access, CRAM obtains both the data and the compression status.

\begin{figure}[htb]
	\centering
    \vspace{-0.13 in}
 \centerline{\epsfig{file=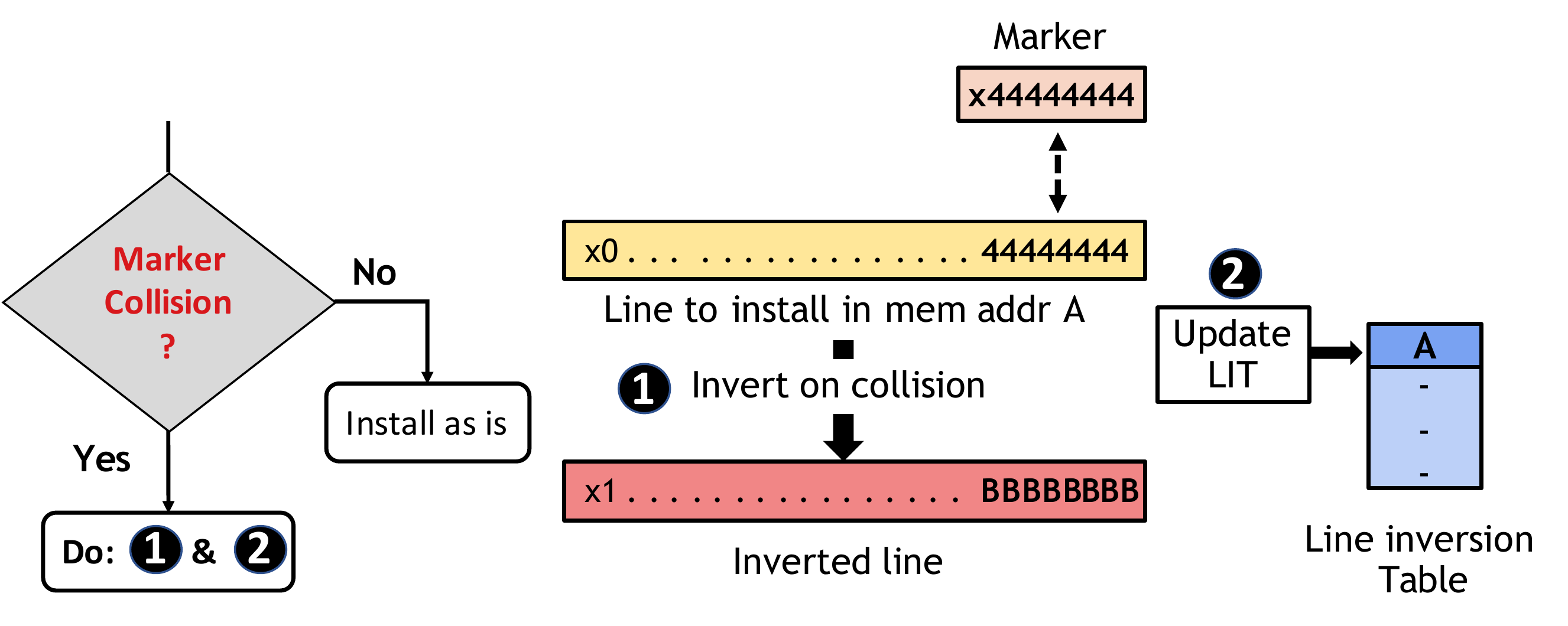, width=\columnwidth}}
	\vspace{-0.08 in}
    \caption{Line inversion handles collisions of uncompressed lines with marker.}
	\vspace{-0.13 in}
	\label{fig:marker_collision}
\end{figure}

\ignore{
\subsubsection{Initializing Marker Values} 

The markers in Figure~\ref{fig:marker} were chosen for simplicity of explanation.  We propose that the two marker values be initialized at boot time using a random number generator, ensuring different markers for each machine.  We check that the randomly generated markers are not the complement of each other (required for handling marker collisions).  CRAM needs two registers of 4-bytes each to store the two marker values.  

Furthermore, we can create private "per-line" marker values by hashing the marker with the line address (for example, computing an xor of the marker with the randomized line address, or using a hardware-efficient static randomizer such as Feistel Network~\cite{qureshi:micro09}).  Such per-line marker values ensure that the likelihood of uncompressed data colliding with the marker value for a large number of lines remains negligibly small, even if the program writes identical values to a large portion of memory. Note that the per-line markers get computed at runtime based on the two global markers, so the storage required for having per-line marker remains a total of 8 bytes (2 registers of 4 bytes each) for the entire memory.

}
\label{label:collisions}
\vspace{0.05 in}
\noindent{\bf Handling Collisions with Marker via Inversion:}  We define a {\em marker collision} as the scenario where the data in an uncompressed line (last four bytes) matches with one of the markers. Since our design generates per-line markers, the likelihood of marker collision is quite rare (less than one in a billion). However, we still need a way to handle it without incurring significant storage or complexity.  CRAM handles marker collisions simply by inverting the uncompressed line and writing this inverted line to memory, as shown in Figure~\ref{fig:marker_collision}. Doing so ensures that the only lines in memory that contain the marker value (in the last four bytes) are compressed lines.

A dedicated on-chip structure, called the {\em Line Inversion Table (LIT)}, keeps track of all the lines in memory that are stored in an inverted form.  The likelihood that multiple lines resident in memory concurrently encounter marker collisions is negligibly small.  
For example, if the system continuously writes to memory, then it will take more than 10 million years to obtain a scenario where more than 16 lines are concurrently stored in inverted form. Therefore, for our 16GB memory, we provision a 16-entry LIT in CRAM.

When a line is fetched from memory, it is not only checked against the marker, but also against the complement of the marker.  If the line matches with the inverted value of the marker, then we know that the line is uncompressed.  However, we do not know if the retrieved data is the original data for the line or if the line was stored in memory in an inverted form due to a collision with the marker.  In such cases, we consult the LIT.  If the line address is present in the LIT, then we know the line was stored in an inverted fashion and we will write the reverted value in the LLC.  Otherwise, the data obtained from the memory is written as-is to the LLC.  

On a write to the memory, if the line address is present in the LIT, and the last four bytes of the line no longer match with any of the markers, then we write the line in its original form and remove this line address from the LIT.  Each entry in the LIT contains a valid bit and the line address (30 bits), so our 16-entry LIT incurs a storage overheads of only 64 bytes.  We recommend that the size of the LIT be increased in proportion to the memory size.

\vspace{0.1 in}

\noindent\textbf{Efficiently Handling LIT Overflows:}  In the extremely rare cases LIT can overflow, and we have two solutions to handle this scenario: (Option-1) Make the LIT memory-mapped (one inversion-bit for every line in memory, stored in memory) and this can support every line in memory having a collision. On marker-collision, the memory system has to make two accesses: one access to the memory, and another to the LIT to resolve collision. Under adversarial settings, the worst-case effect would simply be twice the bandwidth consumption. We implement updates to the LIT by resetting the LIT entry when lines with marker-collisions are brought into the LLC and marking these cachelines as dirty. On eviction, these lines will be forced to go through the marker-collision check and will appropriately set the corresponding LIT entry. (Option-2) On an LIT overflow, CRAM can regenerate new marker values using the random number generator, encode the entire memory with new marker values, and resume the execution.  As cases of LIT overflows are rare (once per 10 million years), the latency of handling LIT overflows does not affect performance.

\ignore{
\footnote{An alternative to re-encoding the entire memory is to make the LIT memory-mapped and have one bit per line, which denotes if the line is inverted or not.  Note that, even though this would consume memory space, LIT is accessed infrequently (less than one in a billion accesses), so the bandwidth overhead of such memory-mapped LIT is negligibly small.  We use the re-encoding based implementation as it avoids having to specify that some portion of the memory (reserved for LIT) is unavailable for use by the OS.} 
}

\ignore{
1. problem with naively picked codes
2. DOS with frequent LIT overflow causing reencryption of mem scheme
3. cryptographically secure marker - Feistel Networks
}

\label{sssec:attack}
\vspace{0.1 in}
\noindent{\bf Attack-Resilient Marker Codes:} The markers in Figure~\ref{fig:marker} were chosen for simplicity of explanation. Markers generated from simple address based hash functions can be a target for a Denial-of-Service Attack. An adversary with knowledge of the hash function can write data values intended to cause frequent LIT overflows resulting in severe performance degradation.  We address this vulnerability by using a cryptographically secure hash function (e.g. such as DES\cite{DES}, given that marker generation can happen off-the-critical path) to generate marker values on a per-line basis. This would make the marker values impractical to guess without knowledge of the secret-keys of the hash function, which are generated randomly for each machine. Furthermore, the secret-keys are regenerated in the event of an LIT overflow which changes the per-line markers.



\begin{figure}[htb]
	\centering
\centerline{\epsfig{file=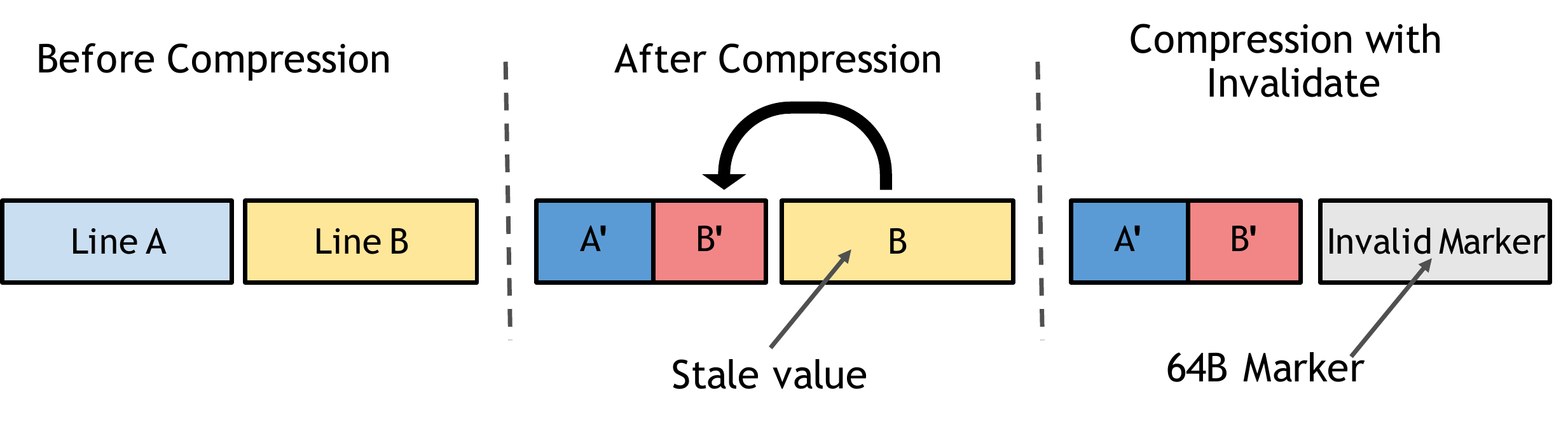, width=\columnwidth}}
	\caption{Compression relocates lines and can create copies of data. We mark such lines as invalid to ensure correct operation.}
	\label{fig:cram_copies}
\end{figure}

\vspace{0.1 in}
\noindent{\bf Efficiently Invalidating Stale Data:}  Compression can relocate the lines, and, when lines get moved, they can leave behind a potentially stale copy of the line. For example, in Figure~\ref{fig:cram_copies}, if adjacent lines A and B became compressible (into values A' and B'), we could move B' and store lines A' and B' together in one physical location. However, an old value of B would still exist in the previous location. Reading the previous location would reveal an old value of B that could still be erroneously interpreted as a valid uncompressed cacheline. Keeping all locations of the line in sync requires significant bandwidth overheads. Therefore, we simply mark such lines as invalid using a special 64-byte marker value, called {\em Invalid Line Marker (Marker-IL)}.  Marker-IL is also initialized at boot time using a randomly generated value. Per-line Marker-IL can be generated as in Section~\ref{sssec:attack}. Collisions with Marker-IL are extremely rare (1 in $2^{512}$ probability, less than one in quadrillion years), and are also handled using line inversion, and are tracked by the LIT.

\begin{figure*}[htb]
	\centering
    \vspace{-.20in}
\centerline{\epsfig{file=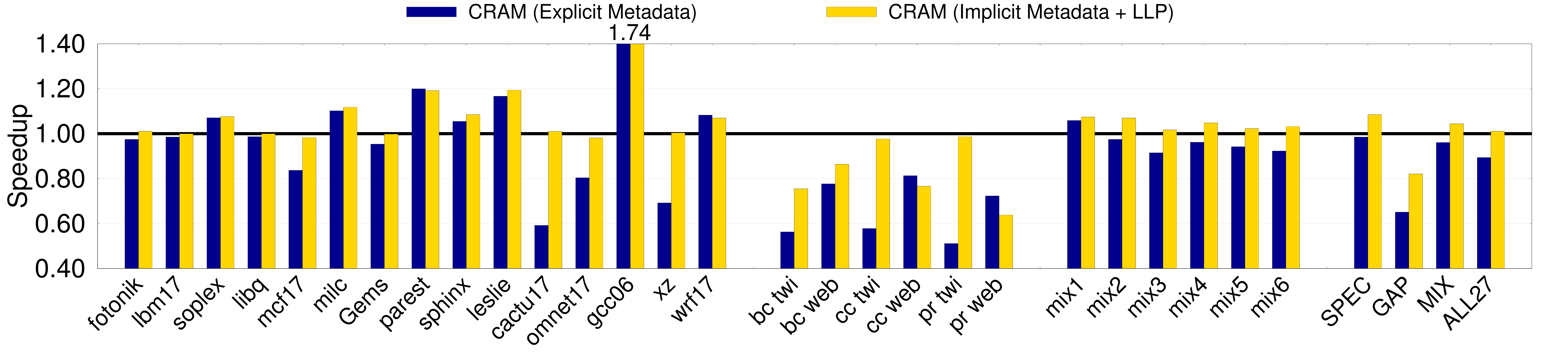, height=1.6 in}}
	\vspace{-0.15 in}
	\caption{Speedup of CRAM with explicit metadata and CRAM with implicit metadata, normalized to uncompressed memory. CRAM with implicit metadata eliminates metadata lookup and improves performance.}
	\vspace{-0.13 in}
	\label{fig:cram_performance}
\end{figure*}



\vspace{0.1 in}

\noindent\textbf{Handling Updates to Compressed Lines:}  An update to a compressed line can render the entire group (of two or four lines) from compressible to incompressible. Such updates must be performed carefully so that the data of the other line(s) in the group gets relocated to their original location(s).  To accomplish this, we need to know the compressibility of the line when the line was obtained from memory.  To track this information, we provision 2-bits in the tag store of the LLC that denotes the compression level when the data was read from memory. On an eviction, we can determine if the lines were previously uncompressed, 2-to-1 compressed, or 4-to-1 compressed by checking these two bits, and we can send writes and invalidates (when applicable) to the appropriate locations.

\vspace{0.1 in}

\noindent\textbf{Ganged Eviction:} Write-back of a cacheline that belongs to a compressed group can require a read-modify-write operation if the other cachelines in the group are not present in the cache. Our design avoids this by using a ganged-eviction scheme which forces the eviction of all members of a compressed group if one of its members gets evicted. This ensures that all the members of a compressed group are either simultaneously present or absent from the LLC, effectively avoiding the need for read-modify-write operations. Our evaluations show that ganged eviction has negligible impact on the LLC hit rate.




\newpage


\subsection{Prediction for Line Location}

With implicit-metadata, CRAM can efficiently determine the compressibility status of any line retrieved from memory.  Reading the line from an incorrect location returns the invalid-line marker (Marker-IL).  However, in such cases, a second request must be sent to another location to obtain the line (for example, an access to B gets routed to A because A contains both A and B).  Sending multiple accesses to retrieve a line from memory wastes bandwidth.  To obtain the line in a single access (in the common case), we develop a {\em Line Location Predictor (LLP)}, that predicts the compressibility status of the line. Knowing the compressibility helps in determining the location of the line (e.g. B will be in original location if incompressible and at A if compressible).

\begin{figure}[htb]
\vspace{-0.10 in}
	\centering
 \centerline{\epsfig{file=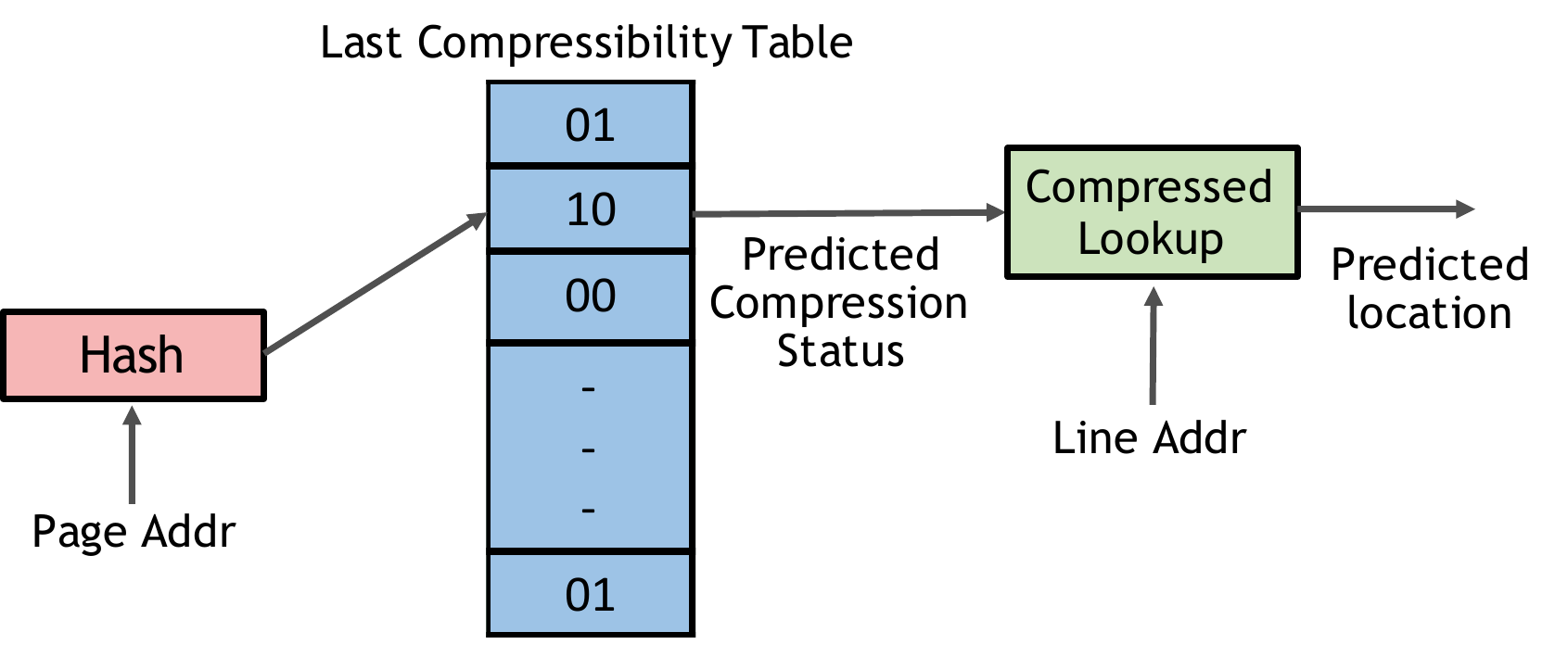, width=3 in}}
	\caption{Line Location Predictor uses line address and compressibility-prediction (based on last-time compressibility) to predict location.}
	\label{fig:predictor}
\end{figure}

\ignore{
To design a low-cost LLP, we exploit the observation that lines within a page are likely to have similar compressibility~\cite{LCP:MICRO2013}\cite{dice}. Cacheline compressibility is tracked at a page granularity using a Page Compression Table (PCT). In order to determine the location of a cacheline, the page address can be used to index into the PCT to predict the compressibility of the line. The compression information along with the line address can be used to predict the location of the line in memory. The LCT is updated on a misprediction so that it tracks the most recently seen compressibility for each page. We use a 512-entry PCT, so the total storage overhead is 128 bytes.}

To design a low-cost LLP, we exploit the observation that lines within a page are likely to have similar compressibility\cite{LCP:MICRO2013}\cite{dice}. Figure~\ref{fig:predictor} shows the organization of the LLP. LLP contains the {\em Last Compressibility Table (LCT)}, that tracks the last compression status seen for a given index. The LCT is indexed with the hash of the page address. So, for a given access, the index corresponding to the page address is used to predict the compressibility, then line location. The LCT is used only when a prediction is needed (for example, A is always resident in its own location and does not need a prediction). We use a 512-entry LCT, so the storage overhead is 128 bytes.

\ignore{
To design a low-cost LLP, we exploit the observation that lines within a page are likely to have similar compressibility~\cite{LCP:MICRO2013}\cite{dice}. Organization of the LLP is shown in Figure~\ref{fig:predictor}. LLP consists of a {\em Page Compressibility Table (PCT)}, that tracks the last compression status seen for a given page. 
For a new access, the 
page address is used to index into the PCT and uses the last compression status seen as the compressibility prediction. The line address along with the predicted compression status is then used to predict the location of the line. We use a 512-entry PCT, so the total storage overhead is 128 bytes.
}


\ignore{
Similar compressibility within page.
We want: last-compressibility seen for a particular physical page.
To get this: we store 

To design a low-cost LLP, we exploit the observation that lines within a page are likely to have similar compressibility~\cite{LCP:MICRO2013}\cite{dice}. The compression status of cachelines are tracked at a page granularity using a Page Compression Table (PCT). In order to determine the location of a cacheline, the page address can be used to index into the PCT to predict the compressibility of the line. The compression information along with the line address can be used to predict the location of the line in memory. The LCT is updated on a misprediction so that it tracks the most recently seen compressibility for each page.

}

\ignore{
To design a low-cost LLP we exploit the observation that lines within a page are likely to have similar compressibility~\cite{LCP:MICRO2013}\cite{dice}. Organization of the LLP is shown in Figure~\ref{fig:predictor}. LLP consists of a {\em Page Compressibility Table (PCT)}, that tracks the last compression status seen for a page mapping to a given index. For a new access, the 
page address is used to index into the PCT to determine the compressibility of the lines in the page. The line address along with the predicted compression status can be used to predict the location of the line within a page. We use a 512-entry PCT, so the total storage overhead is 128 bytes.
We use a 512-entry LCT, so the total storage overhead is 128 bytes.
}

\begin{figure}[htb]
	\centering
 \centerline{\epsfig{file=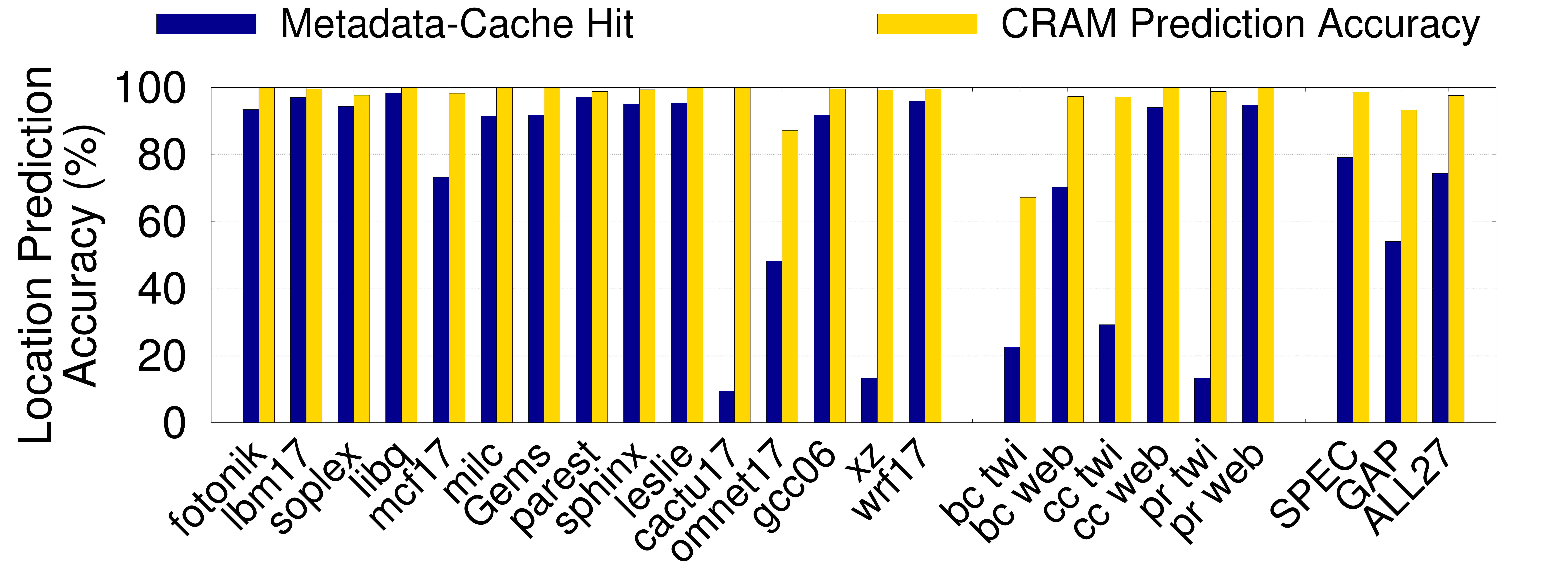, width=\columnwidth}}
     \vspace{-0.15 in}
	\caption{Probability of finding line in one access for explicit-metadata and CRAM with LLP predictor.}
    \vspace{-0.07in}
	\label{fig:predictor_acc}
   \end{figure}

With explicit-metadata, if there is a hit in the metadata cache, we can determine the location of the line and obtain the line is one memory access.  However, a miss in the metadata cache means we need to send two access, one for the metadata and second for the data.   Figure~\ref{fig:predictor_acc} compares the hit-rate of the metadata cache (32KB) with the prediction accuracy of the LLP (128 bytes).  Even though the LLP is quite small, it provides an accuracy of 98\%, much higher than the hit-rate of the metadata cache. On an LLP misprediction, we re-issue the request to the other possible locations of the line.


\subsection{Speedup of CRAM with Optimizations}

CRAM, when combined with implicit-metadata and LLP, can accomplish the task of locating and interpreting lines, without the need for a separate metadata lookup. Figure~\ref{fig:cram_performance} shows the performance of  CRAM (with implicit-metadata + LLP) compared to the basic CRAM design (with explicit-metadata).  CRAM (with implicit-metadata) eliminates the metadata lookup, which significantly helps both compressible and incompressible workloads. For SPEC workloads, CRAM provides a speedup. However, for Graph workloads, CRAM still causes a slowdown.  We investigate bandwidth of CRAM to determine the cause.


\subsection{Bandwidth Breakdown of CRAM}

\begin{figure}[htb]
	\centering
    	\vspace{-0.05 in}
\centerline{\epsfig{file=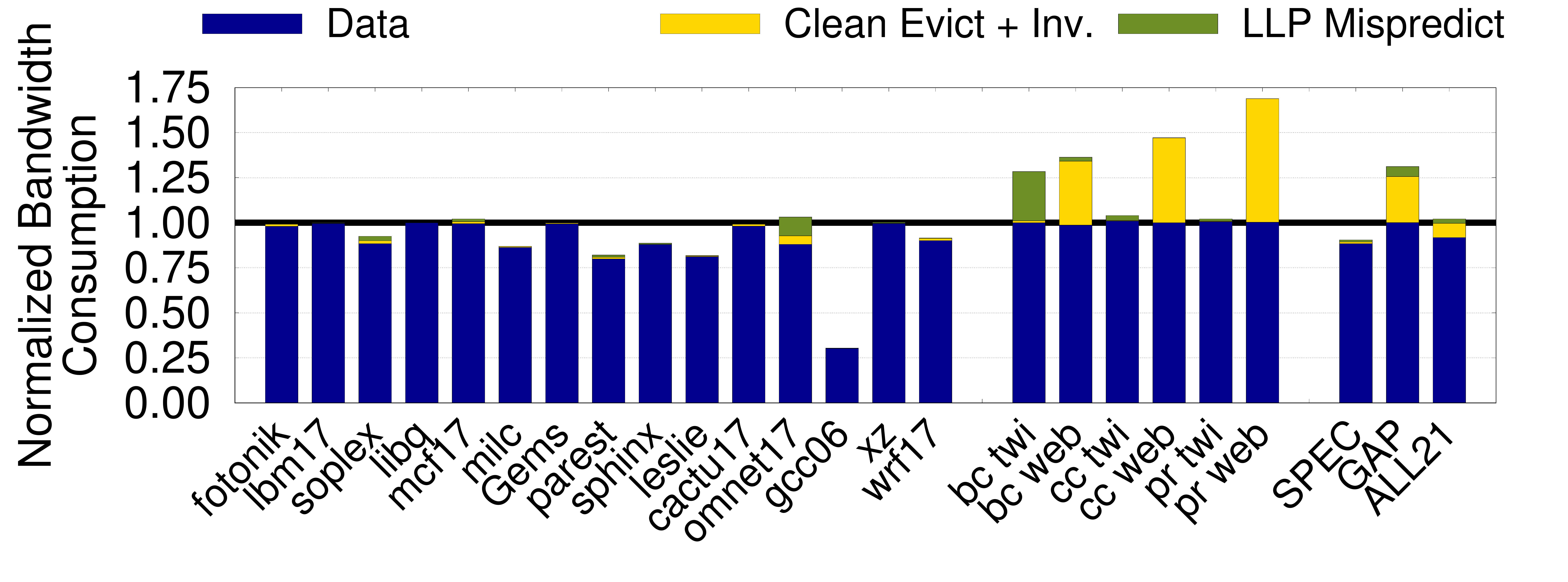, width=\columnwidth}}
	\vspace{-0.15 in}
	\caption{Bandwidth consumption for Optimized CRAM approach, normalized to uncompressed memory.}
	\vspace{-0.17 in}
	\label{fig:cram_bandwidth}
\end{figure}

Figure~\ref{fig:cram_bandwidth} shows the bandwidth consumption of CRAM (with implicit-metadata + LLP), normalized to uncompressed memory.
The components of bandwidth consumption of CRAM are  data, second access due to LLP  mispredictions, and clean writebacks + invalidates (for writing compressed data). High location prediction accuracy means we are able to effectively remove the cost of metadata lookup, except for \textit{bc\_twi}.  For Graph workloads, the inherent cost of compression (i.e., compressing and writing back clean lines, and invalidating) is the dominant source of bandwidth overhead and the cause for performance degradation. We develop an effective scheme to disable compression when compression degrades performance.

\begin{figure*}[htb]
	\vspace{-0.15 in}
	\centering
\centerline{\epsfig{file=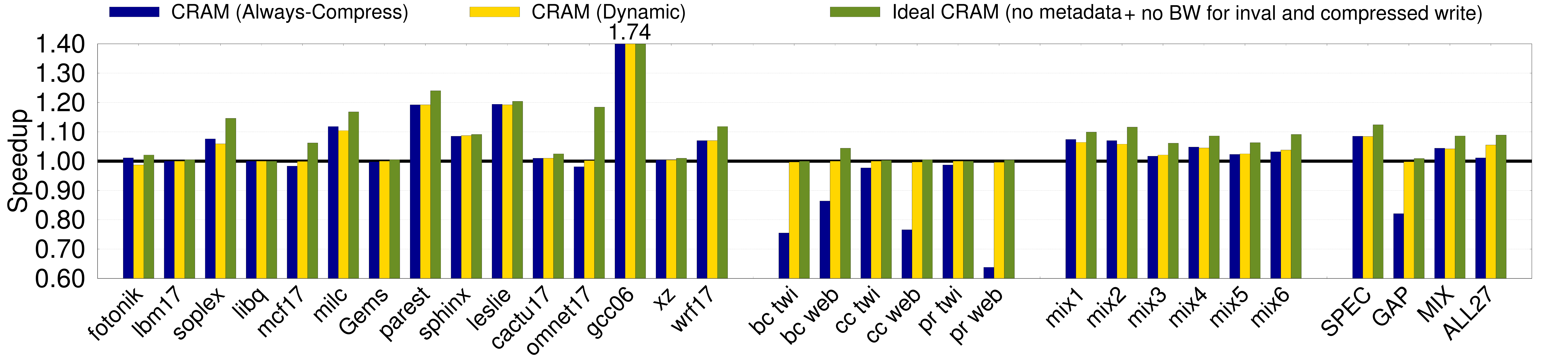, height=1.6 in}}
	\vspace{-0.08 in}
	\caption{Speedup of Static-CRAM, Dynamic-CRAM, and Ideal memory compression. Dynamic-CRAM avoids slowdown for workloads that do not benefit from compression, and performs similar to ideal scheme with no overheads.}
	\vspace{-0.13 in}
	\label{fig:dcram_performance}
\end{figure*}


\section{CRAM: Dynamic Design}
Thus far, we have focused only on avoiding the metadata access overheads of compressed memory.  However, even after eliminating all of the bandwidth overheads of the metadata, there is still performance degradation for several workloads.  Compression requires additional writebacks to memory which can consume additional bandwidth. For example, when a cacheline is found to be compressible on eviction from LLC, it needs to be written back in its compressed form to memory. What could have been a clean evict in an uncompressed memory is now an additional writeback, which becomes a bandwidth overhead.\footnote{CRAM installs new pages in an uncompressed form to avoid inaccurate prefetches. By compressing adjacent lines that are evicted from the LLC, we can ensure that prefetches are done only when the neighboring lines have been previously accessed together and are thus expected to be useful.} Additionally, CRAM requires invalidates to be sent, which further adds to the bandwidth cost of implementing compression. In general, if the workload has enough reuse and spatial locality, the bandwidth cost of compression yields bandwidth savings in the long run. But for a workload with poor reuse and spatial locality (such as several Graph workloads), the cost of compression does not get recovered, causing performance degradation. 

\ignore{
\section{CRAM: Dynamic Design}
\hl{CRAM installs new pages in memory in an uncompressed form. Compression is done only when adjacent cachelines are found to be compressible during eviction from LLC. This ensures that prefetches happen only when it is likely to be useful but unfortunately this also results in additional bandwidth overheads due to compressed writebacks.

For example, when a cacheline is found to be compressible on eviction from LLC, it needs to be written back in its compressed form to memory. What could have been a clean evict in an uncompressed memory is now an additional writeback, which becomes a bandwidth overhead. Additionally, CRAM requires invalidates to be sent, which further adds to the bandwidth cost of implementing compression. In general, if the workload has enough reuse and spatial locality, the bandwidth cost of compression yields bandwidth savings in the long run. But for a workload with poor reuse and spatial locality (such as several Graph workloads), the cost of compression does not get recovered, causing performance degradation.
}
}
\subsection{Design of Dynamic-CRAM}

We can avoid the degradation by dynamically disabling compression, when compression is found to degrade performance. Doing so would return the workload its baseline performance with an uncompressed memory.  We call this design {\em Dynamic-CRAM}.  Dynamic-CRAM compares at runtime the "bandwidth cost of doing compression" with the "bandwidth benefits from compression", and enabling or disabling compression based on this cost-benefit analysis.


\textbf{Bandwidth Cost of Compression:} The bandwidth overhead of compression comes from sending \textit{extra writebacks} (compressed writebacks from clean locations), sending \textit{invalidates}, and sending requests to \textit{mispredicted} locations. These are additional requests incurred due to compression that could have been avoided if we had used an uncompressed design. 


\textbf{Bandwidth Benefits of Compression:} Compression provides bandwidth benefits by enabling bandwidth-free prefetching. On reading a compressed line, adjacent lines get fetched without any extra bandwidth. This saves bandwidth if the prefetched lines are useful. Tracking \textit{useful prefetches} can allow us to determine the benefits from compression.

\begin{figure}[htb]
	\centering
\centerline{\epsfig{file=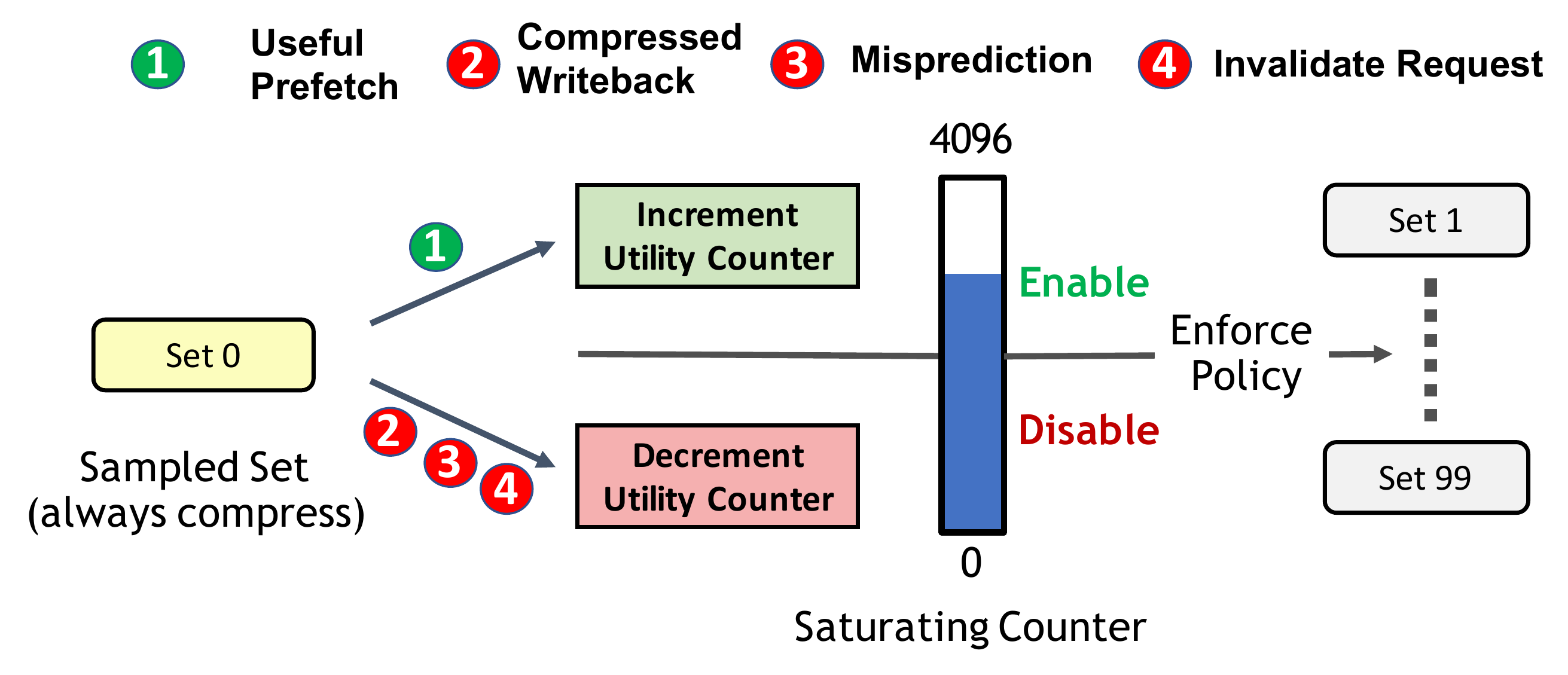, width=\columnwidth}}
	\vspace{-0.05 in}
	\caption{Dynamic-CRAM analyzes cost-benefit of compression on sampled sets. This \ignore{cost-benefit }analysis determines compression policy for the other sets.} 
	\vspace{-0.08 in}
	\label{fig:dcram_operation}
\end{figure}

\ignore{Dynamic-CRAM monitors the bandwidth costs and benefits of compression at run-time, to determine if compression should be enabled or disabled.  To efficiently implement Dynamic-CRAM, we use page-sampling, whereby a small fraction of pages (1\% in our study) always implement compression and we track the cost-benefit statistics only for the sampled pages.  The decision for the remaining (99\%) of the pages is determined by the cost-benefit analysis on the sampled pages, as shown in Figure~\ref{fig:dcram_operation}.  To track the cost and benefit of compression, we use a simple saturating counter. The counter is decremented on seeing the bandwidth \textit{cost} and is incremented on seeing the bandwidth \textit{benefit} of compression.  The Most Significant Bit (MSB) of the counter determines if the compression should be enabled or disabled for the remaining pages.  We use a 12-bit counter in our design. We extend Dynamic-CRAM to support per-core decision by maintaining a 12-bit counter per core.}

Dynamic-CRAM monitors the bandwidth costs and benefits of compression at run-time, to determine if compression should be enabled or disabled.  To efficiently implement Dynamic-CRAM, we use set-sampling, whereby a small fraction of sets in the LLC (1\% in our study) always implement compression and we track the cost-benefit statistics only for the sampled sets.  The decision for the remaining (99\%) of the sets is determined by the cost-benefit analysis on the sampled sets, as shown in Figure~\ref{fig:dcram_operation}.  To track the cost and benefit of compression, we use a simple saturating counter. The counter is decremented on seeing the bandwidth \textit{cost} and is incremented on seeing the bandwidth \textit{benefit} of compression.  The Most Significant Bit (MSB) of the counter determines if the compression should be enabled or disabled for the remaining sets.  We use a 12-bit counter in our design. We extend Dynamic-CRAM to support per-core decision by maintaining a 12-bit counter per core and a 3-bit tag storage for the lines in the sampled sets to identify the core that requested the cacheline.

\vspace{-0.08 in}
\subsection{Effectiveness of Dynamic-CRAM}

Figure~\ref{fig:dcram_performance} shows the performance of Optimized CRAM (that always tries to compress), and Dynamic-CRAM.   CRAM without the Dynamic optimization provides performance improvement for SPEC workloads; however, it degrades performance for GAP workloads.  However, Dynamic-CRAM eliminates all of the degradation, ensuring robust performance -- the design is able to obtain performance when compression is beneficial and avoid degradation when compression is harmful.  On average, Dynamic-CRAM provides 6\% performance improvement, nearing two-thirds of the performance of an idealized compression design that does not incur any bandwidth overheads for implementing compression. Thus, Dynamic-CRAM is a robust and efficient way to implement hardware-based main memory compression.  

\begin{figure*}[htb]
	\centering
	\vspace{-0.20 in}
    \centerline{\epsfig{file=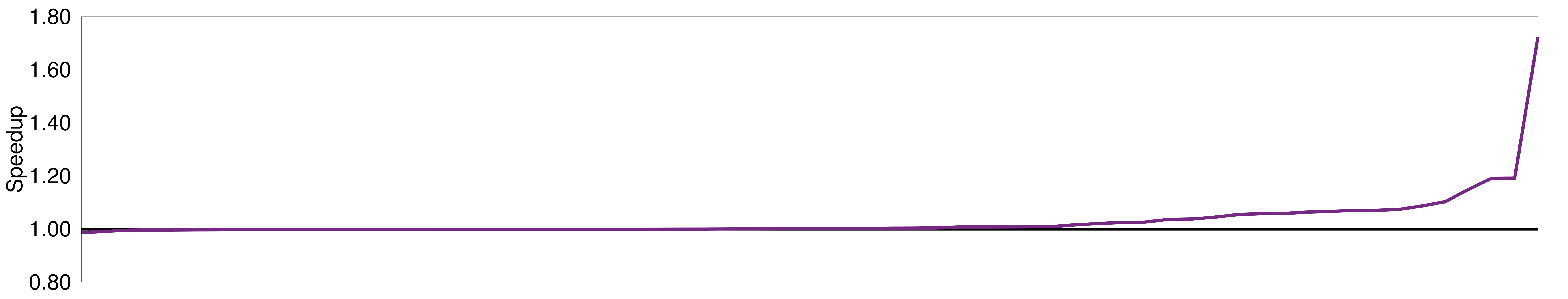, width = \textwidth}}
	\vspace{-0.17 in}
	\caption{S-curve showing speedup of Dynamic-CRAM for 64 workloads, sorted by speedup.} 
	\vspace{-0.1 in}
	\vspace{-0.1 in}
	
	\label{fig:scurve}
\end{figure*}

\newpage


\ignore{DCRAM

\textbf{Inherent overheads of compression}
-Consider an example- clean evictions in an uncompressed scheme do not result in requests to memory. For a compression scheme, clean evicts involve sending extra requests to write back compressed lines. 
But on reading the line again, we would benefit by fetching adjacent lines for no extra bandwidth cost. Our key observation here is that - If the workload has enough data reuse, this bandwidth cost we are paying upfront for compression will yield bandwidth savings in the long run. But for a workload with low data reuse, the cost of compression is never recovered and this leads to performance degradation. We see this problem manifest in the GAP workload which sees a degradation of 30\% with the CRAM design. This clearly violates our goal of providing a robust compression scheme. We need an effective solution to avoid performance degradation for workloads with low data reuse.

A simple way to avoid this degradation is to identify workloads with low data reuse and simply turn off compression. This would return the workload its baseline performance with an uncompressed memory scheme if it does not benefit from compression.

\textbf{measuring usefulness of compression}
In order to do this we need to define a metric to measure usefulness of compression. Compression has cost as well an benefits. We can measure usefulness if we can measure cost and benefit separately.
Cost: Compression involves sending compressed writeback and in case of CRAM, invalidate requests as well. This is extra bandwidth. So, we need to increment our cost measure every time we see a clean compressed writeback or invalidate requests
benefit: Compression provides bandwidth benefits by prefetching adjacent lines at no extra cost. So, useful prefetches are a good way to measure the benefits of compression

By measuring cost and benefits, we have a way to detect when the costs outweigh the benefits and turn off compression to avoid degradation.
We propose a page-sampling scheme to measure this cost-benefit tradeoff and dynamically switch off compression when not beneficial.

page-sampling: We need a sample set to evaluate the cost-benefit tradeoff. We sample 1\% of the pages and always enable compression for these pages. We track the cost and benefit of compression by having a single 12-bit saturating counter. The counter is decremented on seeing the cost and is incremented on seeing the benefit of compression. In CRAM, the cost would constitute clean compressed writebacks or invalidates and the benefits would be useful prefetches for the cachelines that belong to the sampled pages.
The value of the counter would decide if compression should enabled for rest of the lines which are not in the sample set. We simply turn off compression if the value of the counter is below the half-way mark and turn it on if the value is above this mark.
With just a simple 12 bit counter, we can measure the efficacy of the compression scheme and make dynamic decisions to turn off compression when not beneficial. We show that with the Dynamic-CRAM scheme, we avoid degradations in the case of workloads with low data reuse while retaining the benefits for workloads with high data reuse.
}

\ignore{
-We assume that memory is uncompressed to begin with and we check for compressibility with adjacent lines on evictions from L3
	-Figure showing how this works(l3,compression engine,memory)
    -We use ganged compression, compression-groups of 2/4 cachelines
-This changes the location of cachelines based on its compressibility
-We need to know two things to retrieve a cacheline 1. its compression status 2. its location

Lets first look at a naive approach to solving this problem-

Naive Scheme
-One way to solve this is to have a table which stores both these pieces of information \& consult it every time a cacheline needs to be accessed in memory. 
-previous works use a metadata cache to store this information to avoid going to memory every time to retrieve metadata
-Does not work for workloads with high  metadata locality. Could contribute to an overall increase in memory requests. In practice we see degradataions of upto 30\% with this scheme. 

xdram decouples the problem of knowing the compressibility \& location of the cachelines & solves them separately. We first explain how to determine the compression status of cachelines without using a metadata table.

In-lined compressibility information

-we use in-lined markers with compressed cachelines to indicate their compressibility. [Fig]
- this provides a way to determine the compression status of a line read from memory just by examining the marker bits. Obviates need to do metadata lookup.
-We use another marker to indicate invalid memory locations left behind after compression., as shown in fig.
-There is a possibility however that marker bits could collide with an uncompressed cacheline. It would violate correctness if we misinterpreted uncompressed lines as compressed on such collisions.
-We solve this by inverting uncompressed cachelines that collide with the marker at the time they were installed in memory, removing the possibility of collisions with uncompressed lines. The memory controller maintains a small cache of x entries to store memory locations that were inverted
-On reading a line which matches an inverted marker, we need to know how to interpret it.(whether it was inverted on install or if it had the inverted marker bits in its original form). To distinguish between the two, we can consult the metadata cache on inverted marker collision
-If we use a long enough marker, the possibility of this collision is very small. With a 4 Byte marker \& uniformly random data we expect one collision for every x GB of memory.

On writeback, we need to know the prior state of memory so that we can issue invalidates in case the current compression status is different from its prior state. We allocate a 2 bit tag storage in the LLC to indicate the compression status of the line when it was installed in L3.

Location predictor
-Now that we have a way to determine the compressibility of a cacheline read from memory, we need a way to locate the cacheline.
-We propose a compressibility based location predictor to guess the location of the cacheline in memory on a read request.
-Simple last time predictor works with an acccuracy of x\%
-Marker can be used to confirm prediction. Requests can be reissued on mispredict.

}

\ignore{

    \begin{figure}[htb]
	\centering
\centerline{\epsfig{file=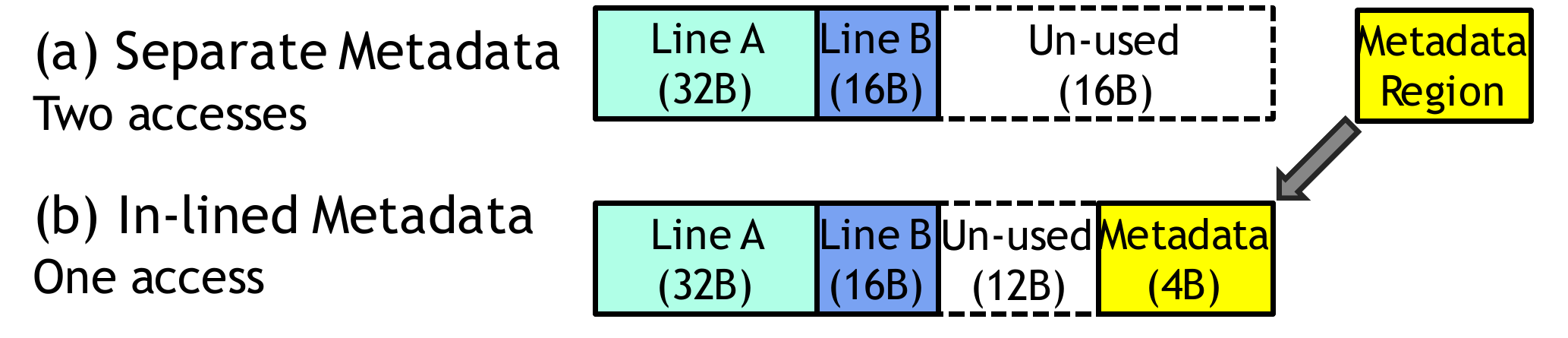, width=\columnwidth}}
	\vspace{-0.08 in}
	\caption{Implicit Metadata using markers: Compressed lines are required to contain the respective markers in the last four bytes.}
	\vspace{-0.13 in}
	\label{fig:motivation_metadata}
\end{figure}

content of the line is written  line address is present in the L, if the 

it actually stores that data value in its original form.  So, in such case not know if the line stores that data

We allocate a portion of the main memory to maintain a Line-Inversion Table (\textit{LIT}) to indicate the lines that are inverted to avoid marker collisions. 
On a read to a line that matches an inverted marker, we consult this table to find if the original data matched the marker or the inverted marker, and invert accordingly to restore the data. The cost of this approach is that we
need to consult this table when we read in a line from memory that matches the inverse of the marker. 

Fortunately, with a 4 Byte marker and uniformly random data, we would expect only one collision (and LIT access) for every 256 GB of memory accessed
Hence, this approach enables us to guarantee that uncompressed lines will not match the marker, and allows us
to interpret any arbitrarily accessed line without the need for a separate metadata lookup.

\textbf{Updating Line Inversion Table:} 
Next, we consider the costs in keeping this inversion table consistent with the data stored in memory. When the data is written to memory, the inversion table needs to be kept up-to-date. We can issue an update to the inversion table every time a line is written, but this update would constitute a significant bandwidth overhead.
We can avoid this bandwidth overhead if we keep track of the prior state of the inversion table and issue updates only if the inversion status is changed. But, this approach would require a storage of 1 bit per line in LLC.


An alternate solution that does not require storage overhead is to reset the inversion bit on every read to an inverted line. This ensures that lines in cache know that the prior inversion state in memory is 0, and will only need to send updates to the inversion table when there is going to be an inverted-marker collision.
However, to ensure that inversion table in memory is kept accurate for previously-inverted lines, inverted lines are installed in the LLC as dirty to force them to keep the inversion table up-to-date on writeback.
This solution enforces updates to the inversion table at no storage cost and negligible bandwidth cost (as inversions are infrequent).
}

%% file: results.tex
\section{Results and Analysis}

\subsection{Storage Overhead of CRAM Structures}

CRAM can be implemented with minor changes at the memory controller.  Table~\ref{table:storage} shows the storage overheads 
\ignore{of the new storage structures } required for implementing Dynamic-CRAM.  The total storage of the additional structures at the memory controller is less than 300 bytes.  In addition to these structures, CRAM needs 2-bits in the tag-store of each line in the LLC to track prior-compressibility. And, per-core Dynamic-CRAM needs 4-bits per each line in sampled sets (1\%) for reuse and core id. 


\begin {table}[ht]
\vspace{-.1in}
\caption{Storage Overhead of CRAM Structures}
\begin{center}
\vspace{-0.13 in}
\begin{small}
\renewcommand{\arraystretch}{.75}
\setlength{\extrarowheight}{2.0pt}{
\begin{tabular}{|c|c|} \hline
Structure               & Storage Cost           \\ \hline \hline
Marker for 2-to-1       & 4 Bytes         \\ \hline
Marker for 4-to-1       & 4 Bytes         \\ \hline
Marker for Invalid Line & 64 Bytes        \\ \hline
Line Inversion Table (LIT)    & 64 Bytes        \\ \hline
Line Location Predictor (LLP) & 128 Bytes       \\ \hline
Dynamic-CRAM counter    & 12 Bytes       \\ \hline \hline
Total                   & 276 bytes \\ \hline
        \end{tabular}
}
\end{small}
\vspace{-0.25 in}
      \label{table:storage}
    \end{center}
\end{table}

\subsection{Extended Evaluation}
\label{ssec:all_workloads}

We perform our study on 27 workloads that are memory intensive. Figure~\ref{fig:scurve} shows the speedup with Dynamic-CRAM across an extended set of 64 workloads (29 SPEC2006, 23 SPEC2017, 6 GAP, and 6 mixes), including ones that are not memory intensive. 
Dynamic-CRAM is robust in terms of performance, as it avoids degradation for any of the workloads while retaining improvement when compression helps.

\ignore{
\subsection{Impact of CRAM on Hit-Rate of L3}

If data is compressible, CRAM can send multiple lines to the L3 with a single access to memory. As these lines are spatially close, these lines are likely to be used within a short period and installing these lines can improve L3 hit-rate. Table~\ref{fig:l3_hitrate} shows l3 hit-rate for an uncompressed memory and a memory using Dynamic-CRAM. On average, CRAM improves hit-rate by XX.X\%.

\begin{table}[hbt]
	\vspace{-.1 in}
	\centering
	\begin{small}
	\setlength{\tabcolsep}{0.085cm}
	\caption{L3 Hit-rate Benefits Under CRAM}
	\vspace{-.05 in}
	\begin{tabular}{|c||c|c|}\hline
		& Uncompressed & Dynamic CRAM\\ \hline \hline
		SPEC  & XX\% & XX\% \\ \hline
		GAP  & XX\% & XX\%  \\ \hline 
        MIX  & XX\% & XX\%  \\ \hline \hline
		ALL27  & XX\% & XX\% \\ \hline
	\end{tabular}
	\label{fig:l3_hitrate}
	\end{small}
	\vspace{-.2 in}
\end{table}
}

\vspace{-.1in}
\subsection{Impact on Energy and Power}

\begin{figure}[htb]
	\centering
	\vspace{-0.08 in}
	\vspace{-0.08 in}
	\centerline{\epsfig{file=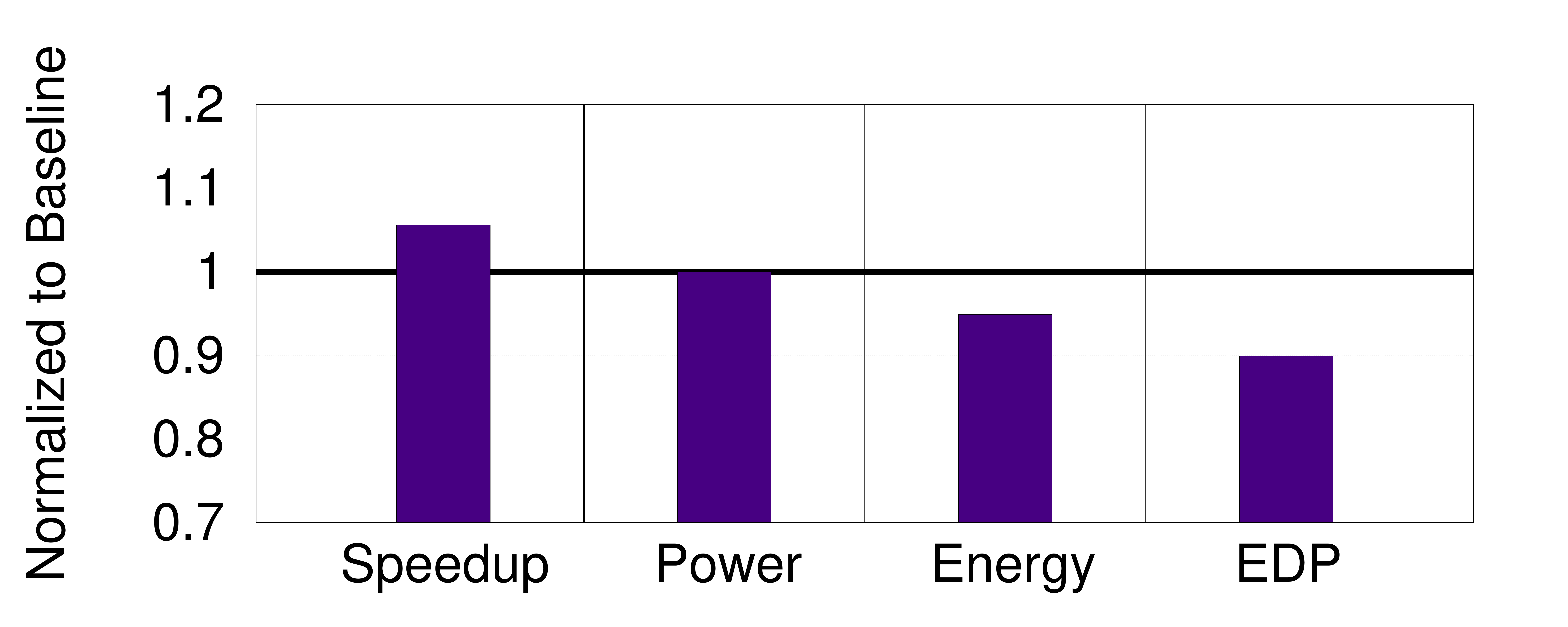, width= 0.9\columnwidth}}
	\vspace{-0.17 in}
    \caption{Dynamic-CRAM impact on energy and power}
	\vspace{-0.1 in}
	\vspace{-0.15 in}
	
	\label{fig:edp}
\end{figure}

Figure~\ref{fig:edp} shows the power, energy consumption and energy-delay-product (EDP) of a system using Dynamic-CRAM, normalized to a baseline uncompressed main memory. Energy consumption is reduced as a consequence of fewer number of requests to main memory. Overall, Dynamic-CRAM reduces energy by 5\% and improves EDP by 10\%.

\subsection{CRAM Sensitivity to Number of Memory Channels}


CRAM offers bandwidth-free adjacent-line prefetch, which are latency benefits that exist regardless of the number of memory channels. Table~\ref{tab:channels} shows that CRAM consistently provides speedup of 5\% even with larger number of channels.

\begin{table}[hbt]
	\vspace{-.13 in}
	\centering
	\caption{CRAM Sensitivity to Number of Memory Channels}
	\vspace{-.10 in}
	\begin{tabular}{|c|c|}\hline
		Num. Channels  & Avg. Speedup of CRAM  \\ \hline \hline
		1   & 4.8\%  \\ \hline
		{\bf 2}   & {\bf 5.5\%} \\ \hline
		4   & 4.6\% \\ \hline
	\end{tabular}
	\label{tab:channels}
	\vspace{-.25 in}
\end{table}

\subsection{Comparison to Larger Fetch for L3}

CRAM can install adjacent lines from the memory to the L3 cache. While this may seem similar to prefetching, we note there is a fundamental difference. CRAM installs additional lines in L3 only when those lines are obtained without any bandwidth overhead. Meanwhile, prefetches result in an extra memory access which incurs additional bandwidth. We compare the performance of next-line prefetching and Dynamic-CRAM Table~\ref{fig:prefetch_perf}. Next-line prefetching causes an average slowdown of 10\%, while CRAM achieves a speedup of 6\% as it obtains adjacent lines without the bandwidth cost.

\begin{table}[hbt]
	\vspace{-.175 in}
	\centering
	\begin{small}
	\setlength{\tabcolsep}{0.085cm}
	\caption{Comparison of CRAM to Next-Line Prefetch}
	\vspace{-.12 in}
	\begin{tabular}{|c||c|c|}\hline
		& Next-Line Prefetch & Dynamic-CRAM\\ \hline \hline
		SPEC  & -5.7\% & +8.5\% \\ \hline
		GAP  & -21.1\% & +0.0\%  \\ \hline 
        MIX  & -7.3\% & +4.2\%  \\ \hline \hline
		ALL27  & -9.7\% & +5.5\% \\ \hline
	\end{tabular}
	\label{fig:prefetch_perf}
	\end{small}
	\vspace{-.25 in}
\end{table}

%% file: related.tex
\begin{figure*}[htb]
\vspace{-.20in}
	\centering
\centerline{\epsfig{file=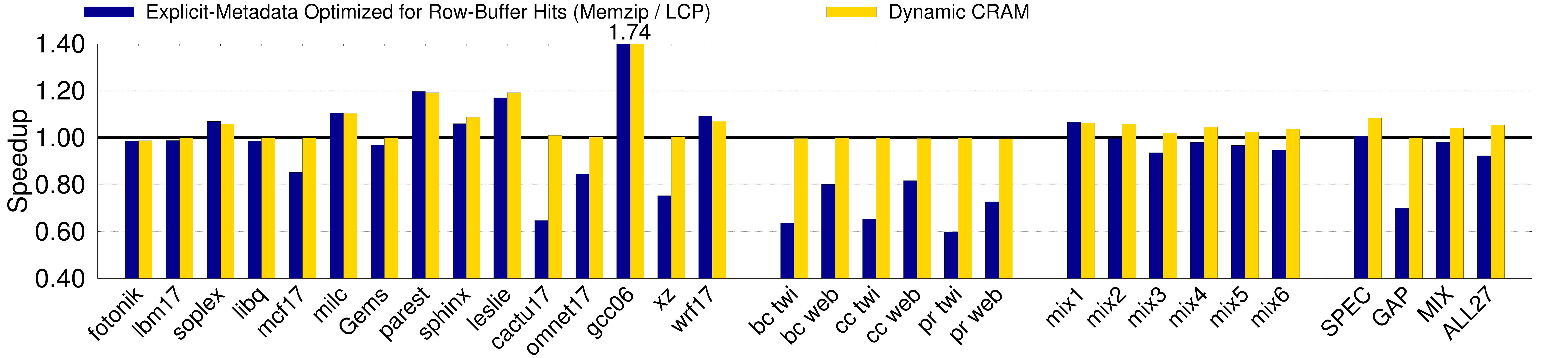, height=1.6 in}}
	\vspace{-0.14 in}
	\caption{Performance of schemes that need explicit metadata management (Memzip, LCP) optimized for row-buffer hits. Prior approaches still require significant bandwidth to retrieve and update metadata.}
	\vspace{-0.20 in}
	\label{fig:memzip_performance}
\end{figure*}

\section{Related Work}

To the best of our knowledge, this is the first paper to propose a robust hardware-based main-memory compression for bandwidth improvement, without requiring any OS-support and without causing changes to the memory organization and protocols. We discuss prior research related to our study.

\vspace{-.1in}
\subsection{Low-Latency Compression Algorithms}

As decompression latency is in the critical path of memory accesses, hardware compression techniques typically use simple per-line compression schemes\cite{FPC}\cite{zca}\cite{zhang2000frequent}\cite{fvc}\cite{BDI}\cite{Alaa:ISCA2004}\cite{bitplane}. We evaluate CRAM using a hybrid compression using FPC~\cite{FPC} and BDI~\cite{BDI}. However, CRAM is orthogonal to the compression algorithm and can be implemented with any compression algorithm, including dictionary-based\cite{cpack}\cite{MORC:MICRO2015}\cite{sc2_huffman}\cite{hycomp}\cite{llc_dedup}.

\vspace{-.1in}
\subsection{Main Memory Compression}

Hardware-based memory compression has been applied to increase the capacity of main memory\cite{IBM-MXT:HPCA2001}\cite{Strenstorm:ISCA2005}\cite{LCP:MICRO2013}. To locate the line, these approaches extend the page table entries to include information on the compressibility of the page. These approaches are attractive as they allow locating and interpreting lines using the TLB. However, such approaches inherently require software-support (from the OS or hypervisor) that limit their applicability.  We want a design that can be built entirely in hardware, without any OS support.

Several studies\cite{shafiee2014memzip}\cite{link_compression}\cite{link_dropwords} propose to send compressed data across links in smaller bursts, and send additional ECC or metadata bits when there is still room in a burst length. These proposals try to improve the bandwidth of the memory system by sending fewer bursts per memory request. However, these proposals require either non-traditional data organization (such as MiniRank)\cite{rank_subsetting1}\cite{ rank_subsetting2}\cite{rank_subsetting3} or changes to the bus protocols or both. CRAM enables compressed memory systems with existing memory devices and protocols.



Prior studies~\cite{LCP:MICRO2013}\cite{shafiee2014memzip} have advocated reducing the latency for metadata lookups by placing the metadata in the same row buffer as the data line. However, this does not reduce the bandwidth required to obtain the metadata.  For comparison, we implement an explicit-metadata scheme optimized to access the same row as the data line.  Figure~\ref{fig:memzip_performance} compares the performance of Dynamic-CRAM with optimized explicit-metadata
provisioned with a 32KB metadata cache. The bandwidth overheads of obtaining metadata is still significant, causing slowdown.  Whereas, Dynamic-CRAM provides performance improvement. 


COP\cite{COP} proposes in-lining ECC into compressed lines and uses the ECC as markers to identify compressed lines. Unfortunately, COP is designed to provide reliability at low cost and provides no performance benefit if the system does not need ECC or already has an ECC-DIMM. Whereas, CRAM is designed to provide bandwidth benefits by fetching multiple lines, and helps regardless of whether the system has ECC-DIMM or not. Furthermore, COP relies on a fairly complex mechanism to handle marker collisions (locking lines in the LLC, memory-mapped linked-list etc.), whereas, CRAM handles marker collisions efficiently via data inversion.





\ignore{
are based on SRAM cache compression,
neglect performance of DRAM cache and need
are tailored for PCM and neglect
 and needs double access for getting associativity or compressibility information. We show how to compress DRAM for bandwidth, without breaking tag access protocol to keep access latency low and without OS support.
}

\vspace{-.1in}
\subsection{SRAM-Cache Compression}
\label{subsection:superblock}

Prior work has looked at using compression to increase capacity of on-chip SRAM caches.  
Cache compression is typically done by accommodating more ways in a cache set and statically allocating more tags~\cite{Alaa:ISCA2004}\cite{base-victim}. 
Recent proposals, such as SCC, investigate reducing SRAM tag overhead by sharing tags across \ignore{spatially-}contiguous sets\ignore{ in what are called {\em superblocks}}\cite{DWood:DecoupledCompressedCache}\cite{DWood:SkewedCompressedCache}\cite{dictionary_sharing}. Compressed caches typically obtain compression metadata by storing metadata beside tag and retrieving them along with tag accesses.  However, these approaches do not scale for memory, as there is no tag space or tag lookup to enable easy access to metadata.

Our restricted data-mapping in CRAM is inspired by the placement in SCC~\cite{DWood:SkewedCompressedCache}, in that the location of the line gets determined by compressibility.  However, unlike SCC, our placement ensures that a significant fraction of lines do not change their locations, regardless of their compression status.  Furthermore, SCC requires skewed-associative lookup of all possible positions, which is possible to do in a cache; however, such unrestricted probes of all possible placement locations would incur intolerable bandwidth overheads in main memory.

\vspace{-.1in}
\subsection{Adaptive Cache-Compression}
\label{subsection:adapcomp}

Prior works have looked at adaptive or dynamic cache compression~\cite{Alaa:ISCA2004}\cite{base-victim}\cite{thread_aware_cache_compression}\cite{DMC} to avoid performance degradation due to latency overheads of decompression or due to extra misses caused by sub-optimal replacement in compressed caches.  These designs are primarily target cache hit rate. Whereas, our main memory proposal targets bandwidth overheads inherent in memory compression (metadata or compressed writes). 
Additionally, fine-grain adaptive memory compression has been previously unexplored, as prior approaches have had no capability to turn off (except by expensive global operation).

\vspace{-.1in}
\subsection{Predicting Cache Indices}
\ignore{
Several studies have looked at predicting indices in associative caches~\cite{hash-rehash}\cite{agarwal_ca}\cite{way_pscache}\cite{way_mru1}\cite{way_mru2}\cite{way_mru3}\cite{way_mru4}\cite{dice}. A cache can verify such predictions simply by checking the tag, and issuing a second request in case of a misprediction. Our work is quite different from these, in that we try to predict the location for \textit{memory}, and memory does not contain any tag information to verify our location prediction (a memory access simply returns 64-bytes of data, without any location information). Our work solves the problem of verifying memory location prediction by integrating implicit-metadata within the line, so a memory access also provides information about whether the location contains compressed data or not.  Our predictors utilize this implicit-metadata to verify the location prediction and issue a request to an alternate location on a misprediction.
}

Several studies have looked at predicting indices in associative caches\cite{hash-rehash}\cite{agarwal_ca}\cite{way_pscache}\cite{way_mru1}\cite{way_mru2}\cite{way_mru3}\cite{way_mru4}\cite{7753256}\cite{dice}. A cache can verify such predictions simply by checking the tag, and issuing a second request in case of a misprediction. Our work is quite different from these, in that we try to predict the location for \textit{memory}. Since memory does not provide tags to identify the data like caches, we verify our prediction by integrating implicit-metadata within the line, allowing memory accesses to provide information about whether the location contains compressed data or not.  Our predictors utilize this implicit-metadata to verify the location prediction and issue a request to an alternate location on a misprediction.




\ignore{
DICE uses multiple indexing schemes (TSI and BAI) in order to get bandwidth benefits of spatial indexing and avoid slowdown when data is incompressible.  Prior work in designing direct-mapped L1 caches have also looked at using multiple indexing schemes in order to reduce conflict misses.  On a miss, these designs~\cite{agarwal_ca, agarwal_hash} check an alternative location (a faraway set in the cache) to find the conflicting line.  Unfortunately, such a design that always requires a second access in case of a cache miss would incur high latency (on hits in second location) and high bandwidth (from extra accesses due to a second lookup on a miss).  Schemes that rely on \ignore{concurrently }looking up multiple locations in parallel~\cite{seznec_skew,zcache} to reduce conflict misses would reduce latency overheads but would incur significant bandwidth overheads for DRAM cache.  Unlike these proposals, DICE avoids the latency and bandwidth of second lookup via index prediction and exploiting properties of a DRAM cache.  Furthermore, the multiple indexing schemes in DICE are not aimed at reducing conflict miss but for increasing cache bandwidth for compressed lines.
}


\ignore{
1alameldeen Adaptive Cache Compression
    L1 L2 compression. Compress for hit rate. don't compress, for latency.
    Decoupled-variable-segment cache
        32 8-byte segments = 4 full lines. but 8 way associative.
    
1alameldeen Interaction of prefetching and compression
    Helps. Can improve more if careful about prefetches.
    Alaa R. Alameldeen, David A. Wood:
Interactions Between Compression and Prefetching in Chip Multiprocessors. HPCA 2007: 228-239
    
1hallnor reinhard. indirect indexed cache    
    fully associative everywhere
    software managed cache

Princeton TORC
    compress across many lines (like deduplicated cache). Increases effective capacity a lot. But increases latency. Would not help DCache.

1 Dwood decoupledmicro13
    first step. look up all tags. ..... nope. Even 1 extra access to get all metadata is pure overhead.

    uses superblocks (quads)(64B*4 ). saves on tag space for neighbor lines. add pointer to use more space
    
    bad: lot of tag checks. must 
    
    superblock tag
    decoupled sectored (or superblock), to increase effective number tags. 
    
    1. normal superblocks have set locations. 
    2. decoupling, by including pointers, reduces fragmentation and more use of cache.

    Superblock: 
        normal: 1 index -> 8 way
        quads: 4 index -> 32 way... or
            1 superblock. with 4 spatially-contiguous lines compressible. share tags. compress to set amounts to reduce search overhead (8B blocks). 
            
    CPACK+Z 9 cycle decompress ratio 3.9 average compression ratio

1 Dwood skewed compressedmicro14
    first step. look up all tags. skewed, so 2? still pure overhead
    superblock tags (similar to shared tags)
    8 way. compressed indexing is skewed. reduces metadata requirement for location. 
        skewed, 
    compress contiguous lines together. similar compressibility.
    
}

\ignore{ 
These are copy pasted or small notes

~\cite{LCP:MICRO2013}

MXT
    2 serial accesses.memory look-up to find pointer to compressed block.  also needs software support to know when physical space full.

Ekman Strenstrom
    The work of Ekman and Stenstrom [16] addresses many
of the weaknesses found in the MXT design. It associates
the metadata for a page along with the TLB entry. This
metadata tracks the size of each compressed block within
the page. In parallel with the LLC look-up, the location of
the cache line in memory is computed based on the infor-
mation in the metadata. This hides the latency for address
calculation, but increases the energy overhead. On every
write, if the new compressed block has a very different size
from the old compressed block, the blocks may have to be
re-organized, requiring a page table update and even requir
-
ing a copy to a new page in the worst case. The proposal
has no explicit feature for bandwidth or energy efficienc
    Store metadata in TLB to reduce 2 serial access

LCP
    compress to small size. similar goal. but impractical in DRAM cache as needs OS support. and handles incompressible lines in high latency.

Memzip
    Compress in main memory (same location). use extra space for ECC
        Link compression - Send data in smaller burst.
        Store size metadata in TLB. Can miss.

QUOTED from memzip -- do not use
    here are three primary hardware-based memory com-
pression architectures for chip multiprocessors (CMPs) an
d
DDR3 memory in recent literature. The IBM MXT tech-
nology [4] uses a memory look-up to find a pointer to a
compressed block. Thus, every access requires two mem-
ory fetches. The work of Ekman and Stenstrom [16] stores
metadata with every TLB entry so that the start of a cache
line can be computed. The LCP architecture [26] optimizes
the common case. For cache lines that can be compressed
within a givensize, a pointerto the compressedblockis triv
-
ially computed. But a cache line that cannot be compressed
within the specified size will require three memory accesses
in the worst case. More details about these schemes are pro-
vided in Section 2.

}

%% file: summary.tex
\vspace{-.1in}
\section{Conclusions}

This paper investigates practical designs for main-memory compression to obtain higher memory bandwidth.  The proposed design, {\em CRAM}, is hardware-based, does not require any OS/hypervisor support, or changes to the memory modules or access protocols.  We show that for compressed memory designs, the bandwidth overheads of accessing metadata can be significant enough to cause slowdown for several workloads.  We propose the implicit-metadata design, based on {\em marker} values, to eliminate the storage and bandwidth overheads of the metadata access. We also propose a simple and effective predictor to predict the location of the line in compressed memory, and a dynamic scheme to disable compression when compression degrades performance. Our proposed design provides an average speedup of 6\%, and avoids slowdown for any of the workloads.  This design can be implemented with minor additions to the memory controller.

%% file: ms.bbl
\begin{thebibliography}{10}
\providecommand{\url}[1]{#1}
\csname url@samestyle\endcsname
\providecommand{\newblock}{\relax}
\providecommand{\bibinfo}[2]{#2}
\providecommand{\BIBentrySTDinterwordspacing}{\spaceskip=0pt\relax}
\providecommand{\BIBentryALTinterwordstretchfactor}{4}
\providecommand{\BIBentryALTinterwordspacing}{\spaceskip=\fontdimen2\font plus
\BIBentryALTinterwordstretchfactor\fontdimen3\font minus
  \fontdimen4\font\relax}
\providecommand{\BIBforeignlanguage}[2]{{%
\expandafter\ifx\csname l@#1\endcsname\relax
\typeout{** WARNING: IEEEtran.bst: No hyphenation pattern has been}%
\typeout{** loaded for the language `#1'. Using the pattern for}%
\typeout{** the default language instead.}%
\else
\language=\csname l@#1\endcsname
\fi
#2}}
\providecommand{\BIBdecl}{\relax}
\BIBdecl

\bibitem{IBM-MXT:HPCA2001}
B.~Abali, H.~Franke, X.~Shen, D.~Poff, and T.~Smith, ``Performance of hardware
  compressed main memory,'' in \emph{High-Performance Computer Architecture,
  2001. HPCA. The Seventh International Symposium on}, 2001.

\bibitem{Strenstorm:ISCA2005}
M.~Ekman and P.~Stenstrom, ``A robust main-memory compression scheme,'' in
  \emph{ACM SIGARCH Computer Architecture News}, vol.~33, no.~2.\hskip 1em plus
  0.5em minus 0.4em\relax IEEE Computer Society, 2005.

\bibitem{LCP:MICRO2013}
\BIBentryALTinterwordspacing
G.~Pekhimenko, V.~Seshadri, Y.~Kim, H.~Xin, O.~Mutlu, P.~B. Gibbons, M.~A.
  Kozuch, and T.~C. Mowry, ``Linearly compressed pages: A low-complexity,
  low-latency main memory compression framework,'' in \emph{Proceedings of the
  46th Annual IEEE/ACM International Symposium on Microarchitecture}, ser.
  MICRO-46.\hskip 1em plus 0.5em minus 0.4em\relax New York, NY, USA: ACM,
  2013. [Online]. Available: \url{http://doi.acm.org/10.1145/2540708.2540724}
\BIBentrySTDinterwordspacing

\bibitem{qualcomm}
Qualcomm, ``Qualcomm centriq 2400 processor,''
  \url{https://www.qualcomm.com/media/documents/files/qualcomm-centriq-2400-processor.pdf},
  2017, [Online].

\bibitem{shafiee2014memzip}
A.~Shafiee, M.~Taassori, R.~Balasubramonian, and A.~Davis, ``Memzip: Exploring
  unconventional benefits from memory compression,'' in \emph{High Performance
  Computer Architecture (HPCA), 2014 IEEE 20th International Symposium
  on}.\hskip 1em plus 0.5em minus 0.4em\relax IEEE, 2014.

\bibitem{FPC}
A.~R. Alameldeen and D.~A. Wood, ``Frequent pattern compression: A
  significance-based compression scheme for l2 caches,'' \emph{Dept. Comp.
  Scie., Univ. Wisconsin-Madison, Tech. Rep}, vol. 1500, 2004.

\bibitem{zca}
\BIBentryALTinterwordspacing
J.~Dusser, T.~Piquet, and A.~Seznec, ``Zero-content augmented caches,'' in
  \emph{Proceedings of the 23rd International Conference on Supercomputing},
  ser. ICS '09.\hskip 1em plus 0.5em minus 0.4em\relax New York, NY, USA: ACM,
  2009. [Online]. Available: \url{http://doi.acm.org/10.1145/1542275.1542288}
\BIBentrySTDinterwordspacing

\bibitem{zhang2000frequent}
Y.~Zhang, J.~Yang, and R.~Gupta, ``Frequent value locality and value-centric
  data cache design,'' in \emph{ACM SIGOPS Operating Systems Review}, vol.~34,
  no.~5.\hskip 1em plus 0.5em minus 0.4em\relax ACM, 2000.

\bibitem{fvc}
\BIBentryALTinterwordspacing
J.~Yang, Y.~Zhang, and R.~Gupta, ``Frequent value compression in data caches,''
  in \emph{Proceedings of the 33rd Annual ACM/IEEE International Symposium on
  Microarchitecture}, ser. MICRO 33.\hskip 1em plus 0.5em minus 0.4em\relax New
  York, NY, USA: ACM, 2000. [Online]. Available:
  \url{http://doi.acm.org/10.1145/360128.360154}
\BIBentrySTDinterwordspacing

\bibitem{BDI}
G.~Pekhimenko, V.~Seshadri, O.~Mutlu, P.~B. Gibbons, M.~A. Kozuch, and T.~C.
  Mowry, ``Base-delta-immediate compression: practical data compression for
  on-chip caches,'' in \emph{Proceedings of the 21st international conference
  on Parallel architectures and compilation techniques}.\hskip 1em plus 0.5em
  minus 0.4em\relax ACM, 2012.

\bibitem{Alaa:ISCA2004}
A.~R. Alameldeen, D.~Wood \emph{et~al.}, ``Adaptive cache compression for
  high-performance processors,'' in \emph{Computer Architecture, 2004.
  Proceedings. 31st Annual International Symposium on}.\hskip 1em plus 0.5em
  minus 0.4em\relax IEEE, 2004.

\bibitem{bitplane}
J.~Kim, M.~Sullivan, E.~Choukse, and M.~Erez, ``Bit-plane compression:
  Transforming delta for better compression in many-core architectures,'' in
  \emph{Computer Architecture (ISCA), 2016 ACM/IEEE 43rd Annual International
  Symposium on}, 2016.

\bibitem{USIMM}
N.~Chatterjee, R.~Balasubramonian, M.~Shevgoor, S.~Pugsley, A.~Udipi,
  A.~Shafiee, K.~Sudan, M.~Awasthi, and Z.~Chishti, ``Usimm: the utah simulated
  memory module,'' \emph{University of Utah, Tech. Rep}, 2012.

\bibitem{pinpoint}
H.~Patil, R.~Cohn, M.~Charney, R.~Kapoor, A.~Sun, and A.~Karunanidhi,
  ``Pinpointing representative portions of large intel itanium programs with
  dynamic instrumentation,'' in \emph{Microarchitecture, 2004. MICRO-37 2004.
  37th International Symposium on}, Dec 2004.

\bibitem{SPEC2006}
\BIBentryALTinterwordspacing
J.~L. Henning, ``Spec cpu2006 benchmark descriptions,'' \emph{SIGARCH Comput.
  Archit. News}, vol.~34, Sep. 2006. [Online]. Available:
  \url{http://doi.acm.org/10.1145/1186736.1186737}
\BIBentrySTDinterwordspacing

\bibitem{spec17}
\BIBentryALTinterwordspacing
S.~P.~E. Corporation, ``Spec cpu® 2017,'' 2017, accessed: 2017-11-10.
  [Online]. Available: \url{https://www.spec.org/cpu2017/}
\BIBentrySTDinterwordspacing

\bibitem{GAP}
\BIBentryALTinterwordspacing
S.~Beamer, K.~Asanovic, and D.~A. Patterson, ``The {GAP} benchmark suite,''
  \emph{CoRR}, vol. abs/1508.03619, 2015. [Online]. Available:
  \url{http://arxiv.org/abs/1508.03619}
\BIBentrySTDinterwordspacing

\bibitem{Davis:matrix2011}
\BIBentryALTinterwordspacing
T.~A. Davis and Y.~Hu, ``The university of florida sparse matrix collection,''
  \emph{ACM Trans. Math. Softw.}, vol.~38, Dec. 2011. [Online]. Available:
  \url{http://doi.acm.org/10.1145/2049662.2049663}
\BIBentrySTDinterwordspacing

\bibitem{DES}
\BIBentryALTinterwordspacing
D.~Coppersmith, ``The data encryption standard (des) and its strength against
  attacks,'' \emph{IBM J. Res. Dev.}, vol.~38, May 1994. [Online]. Available:
  \url{http://dx.doi.org/10.1147/rd.383.0243}
\BIBentrySTDinterwordspacing

\bibitem{dice}
\BIBentryALTinterwordspacing
V.~Young, P.~J. Nair, and M.~K. Qureshi, ``Dice: Compressing dram caches for
  bandwidth and capacity,'' in \emph{Proceedings of the 44th Annual
  International Symposium on Computer Architecture}, ser. ISCA '17.\hskip 1em
  plus 0.5em minus 0.4em\relax New York, NY, USA: ACM, 2017. [Online].
  Available: \url{http://doi.acm.org/10.1145/3079856.3080243}
\BIBentrySTDinterwordspacing

\bibitem{cpack}
X.~Chen, L.~Yang, R.~P. Dick, L.~Shang, and H.~Lekatsas, ``C-pack: A
  high-performance microprocessor cache compression algorithm,'' \emph{Very
  Large Scale Integration (VLSI) Systems, IEEE Transactions on}, vol.~18, 2010.

\bibitem{MORC:MICRO2015}
T.~M. Nguyen and D.~Wentzlaff, ``Morc: A manycore-oriented compressed cache,''
  in \emph{Microarchitecture (MICRO), 2015 48th Annual IEEE/ACM International
  Symposium on}.\hskip 1em plus 0.5em minus 0.4em\relax IEEE, 2015.

\bibitem{sc2_huffman}
A.~Arelakis and P.~Stenstrom, ``Sc2: A statistical compression cache scheme,''
  in \emph{Computer Architecture (ISCA), 2014 ACM/IEEE 41st International
  Symposium on}, June 2014.

\bibitem{hycomp}
\BIBentryALTinterwordspacing
A.~Arelakis, F.~Dahlgren, and P.~Stenstrom, ``Hycomp: A hybrid cache
  compression method for selection of data-type-specific compression methods,''
  in \emph{Proceedings of the 48th International Symposium on
  Microarchitecture}, ser. MICRO-48.\hskip 1em plus 0.5em minus 0.4em\relax New
  York, NY, USA: ACM, 2015. [Online]. Available:
  \url{http://doi.acm.org/10.1145/2830772.2830823}
\BIBentrySTDinterwordspacing

\bibitem{llc_dedup}
\BIBentryALTinterwordspacing
Y.~Tian, S.~M. Khan, D.~A. Jim{\'e}nez, and G.~H. Loh, ``Last-level cache
  deduplication,'' in \emph{Proceedings of the 28th ACM International
  Conference on Supercomputing}, ser. ICS '14.\hskip 1em plus 0.5em minus
  0.4em\relax New York, NY, USA: ACM, 2014. [Online]. Available:
  \url{http://doi.acm.org/10.1145/2597652.2597655}
\BIBentrySTDinterwordspacing

\bibitem{link_compression}
\BIBentryALTinterwordspacing
V.~Sathish, M.~J. Schulte, and N.~S. Kim, ``Lossless and lossy memory i/o link
  compression for improving performance of gpgpu workloads,'' in
  \emph{Proceedings of the 21st International Conference on Parallel
  Architectures and Compilation Techniques}, ser. PACT '12.\hskip 1em plus
  0.5em minus 0.4em\relax New York, NY, USA: ACM, 2012. [Online]. Available:
  \url{http://doi.acm.org/10.1145/2370816.2370864}
\BIBentrySTDinterwordspacing

\bibitem{link_dropwords}
\BIBentryALTinterwordspacing
H.~Kim, P.~Ghoshal, B.~Grot, P.~V. Gratz, and D.~A. Jim{\'e}nez, ``Reducing
  network-on-chip energy consumption through spatial locality speculation,'' in
  \emph{Proceedings of the Fifth ACM/IEEE International Symposium on
  Networks-on-Chip}, ser. NOCS '11.\hskip 1em plus 0.5em minus 0.4em\relax New
  York, NY, USA: ACM, 2011. [Online]. Available:
  \url{http://doi.acm.org/10.1145/1999946.1999983}
\BIBentrySTDinterwordspacing

\bibitem{rank_subsetting1}
H.~Zheng, J.~Lin, Z.~Zhang, E.~Gorbatov, H.~David, and Z.~Zhu, ``Mini-rank:
  Adaptive dram architecture for improving memory power efficiency,'' in
  \emph{2008 41st IEEE/ACM International Symposium on Microarchitecture}, Nov
  2008.

\bibitem{rank_subsetting2}
\BIBentryALTinterwordspacing
D.~H. Yoon, M.~K. Jeong, and M.~Erez, ``Adaptive granularity memory systems: A
  tradeoff between storage efficiency and throughput,'' in \emph{Proceedings of
  the 38th Annual International Symposium on Computer Architecture}, ser. ISCA
  '11.\hskip 1em plus 0.5em minus 0.4em\relax New York, NY, USA: ACM, 2011.
  [Online]. Available: \url{http://doi.acm.org/10.1145/2000064.2000100}
\BIBentrySTDinterwordspacing

\bibitem{rank_subsetting3}
J.~H. Ahn, N.~P. Jouppi, C.~Kozyrakis, J.~Leverich, and R.~S. Schreiber,
  ``Future scaling of processor-memory interfaces,'' in \emph{Proceedings of
  the Conference on High Performance Computing Networking, Storage and
  Analysis}, Nov 2009.

\bibitem{COP}
D.~J. Palframan, N.~S. Kim, and M.~H. Lipasti, ``Cop: To compress and protect
  main memory,'' in \emph{2015 ACM/IEEE 42nd Annual International Symposium on
  Computer Architecture (ISCA)}, June 2015.

\bibitem{base-victim}
J.~Guar, A.~R. Alameldeen, and S.~Subramoney, ``Base-victim compression: An
  opportunistic cache compression architecture,'' in \emph{Computer
  Architecture (ISCA), 2016 ACM/IEEE 43rd Annual International Symposium on},
  2016.

\bibitem{DWood:DecoupledCompressedCache}
S.~Sardashti and D.~A. Wood, ``Decoupled compressed cache: Exploiting spatial
  locality for energy-optimized compressed caching,'' in \emph{Proceedings of
  the 46th Annual IEEE/ACM International Symposium on Microarchitecture}.\hskip
  1em plus 0.5em minus 0.4em\relax ACM, 2013.

\bibitem{DWood:SkewedCompressedCache}
S.~Sardashti, A.~Seznec, D.~Wood \emph{et~al.}, ``Skewed compressed caches,''
  in \emph{Microarchitecture (MICRO), 2014 47th Annual IEEE/ACM International
  Symposium on}.\hskip 1em plus 0.5em minus 0.4em\relax IEEE, 2014.

\bibitem{dictionary_sharing}
B.~Panda and A.~Seznec, ``Dictionary sharing: An efficient cache compression
  scheme for compressed caches,'' in \emph{49th Annual IEEE/ACM International
  Symposium on Microarchitecture, 2016}, 2016.

\bibitem{thread_aware_cache_compression}
Y.~Xie and G.~H. Loh, ``Thread-aware dynamic shared cache compression in
  multi-core processors,'' in \emph{2011 IEEE 29th International Conference on
  Computer Design (ICCD)}, Oct 2011.

\bibitem{DMC}
S.~Kim, S.~Lee, T.~Kim, and J.~Huh, ``Transparent dual memory compression
  architecture,'' in \emph{2017 26th International Conference on Parallel
  Architectures and Compilation Techniques (PACT)}, Sept 2017.

\bibitem{hash-rehash}
\BIBentryALTinterwordspacing
A.~Agarwal, J.~Hennessy, and M.~Horowitz, ``Cache performance of operating
  system and multiprogramming workloads,'' \emph{ACM Trans. Comput. Syst.},
  vol.~6, Nov. 1988. [Online]. Available:
  \url{http://doi.acm.org/10.1145/48012.48037}
\BIBentrySTDinterwordspacing

\bibitem{agarwal_ca}
A.~Agarwal and S.~D. Pudar, \emph{Column-associative caches: A technique for
  reducing the miss rate of direct-mapped caches}.\hskip 1em plus 0.5em minus
  0.4em\relax ACM, 1993, vol.~21, no.~2.

\bibitem{way_pscache}
\BIBentryALTinterwordspacing
J.~J. Valls, A.~Ros, J.~Sahuquillo, and M.~E. Gomez, ``Ps-cache: An
  energy-efficient cache design for chip multiprocessors,'' \emph{J.
  Supercomput.}, vol.~71, Jan. 2015. [Online]. Available:
  \url{http://dx.doi.org/10.1007/s11227-014-1288-5}
\BIBentrySTDinterwordspacing

\bibitem{way_mru1}
\BIBentryALTinterwordspacing
B.~Calder, D.~Grunwald, and J.~Emer, ``Predictive sequential associative
  cache,'' in \emph{Proceedings of the 2Nd IEEE Symposium on High-Performance
  Computer Architecture}, ser. HPCA '96.\hskip 1em plus 0.5em minus 0.4em\relax
  Washington, DC, USA: IEEE Computer Society, 1996. [Online]. Available:
  \url{http://dl.acm.org/citation.cfm?id=525424.822662}
\BIBentrySTDinterwordspacing

\bibitem{way_mru2}
D.~H. Albonesi, ``Selective cache ways: On-demand cache resource allocation,''
  in \emph{Microarchitecture, 1999. MICRO-32. Proceedings. 32nd Annual
  International Symposium on}.\hskip 1em plus 0.5em minus 0.4em\relax IEEE,
  1999.

\bibitem{way_mru3}
\BIBentryALTinterwordspacing
M.~D. Powell, A.~Agarwal, T.~N. Vijaykumar, B.~Falsafi, and K.~Roy, ``Reducing
  set-associative cache energy via way-prediction and selective
  direct-mapping,'' in \emph{Proceedings of the 34th Annual ACM/IEEE
  International Symposium on Microarchitecture}, ser. MICRO 34.\hskip 1em plus
  0.5em minus 0.4em\relax Washington, DC, USA: IEEE Computer Society, 2001.
  [Online]. Available: \url{http://dl.acm.org/citation.cfm?id=563998.564007}
\BIBentrySTDinterwordspacing

\bibitem{way_mru4}
H.-C. Chen and J.-S. Chiang, ``Low-power way-predicting cache using valid-bit
  pre-decision for parallel architectures,'' in \emph{19th International
  Conference on Advanced Information Networking and Applications (AINA'05)
  Volume 1 (AINA papers)}, vol.~2, March 2005.

\bibitem{7753256}
A.~Deb, P.~Faraboschi, A.~Shafiee, N.~Muralimanohar, R.~Balasubramonian, and
  R.~Schreiber, ``Enabling technologies for memory compression: Metadata,
  mapping, and prediction,'' in \emph{2016 IEEE 34th International Conference
  on Computer Design (ICCD)}, Oct 2016.

\end{thebibliography}
